\documentclass[a4paper, amsfonts, amssymb, amsmath, reprint, showkeys, nofootinbib]{revtex4-1}
\usepackage[english]{babel}
\usepackage[utf8]{inputenc}
\usepackage[colorinlistoftodos, color=green!40, prependcaption]{todonotes}
\usepackage{amsthm}
\usepackage{mathtools}
\usepackage{physics}
\usepackage{xcolor}
\usepackage{graphicx}
\usepackage[left=23mm,right=13mm,top=35mm,columnsep=15pt]{geometry} 
\usepackage{adjustbox}
\usepackage{placeins}
\usepackage{subfigure}
\usepackage[T1]{fontenc}
\usepackage{lipsum}
\usepackage{csquotes}
\usepackage[pdftex, pdftitle={Article}, pdfauthor={Author},colorlinks,citecolor=blue,linkcolor=blue,urlcolor=blue]{hyperref} 

\def\no{\notag}
\def\veck{\mathbf{k}}
\def\ef{E_{_F}}
\def\phdag{{\phantom \dagger}}

\begin{document}
\title{Spectral functions across an Insulator to Superconductor Transition}

\author{Tamaghna Hazra, Nandini Trivedi, Mohit Randeria}
\affiliation{Department of Physics, Ohio State University, Columbus 43201, USA }

\begin{abstract}
In a minimal 2-band model with attractive interactions between fermions, we calculate the gap to single and two-particle excitations, the band-dependent spectral functions, the superfluid density and compressibility using quantum Monte Carlo (QMC) methods. We find Fermi and Bose insulating phases with signatures of incipient pairing evident in the single-particle spectral functions, and a superconducting state with three different spectral functions: (i) both bands show ``BCS" behavior in which the minimum gap locus occurs on a closed contour on the underlying Fermi surface; (ii) both bands show ``BEC" behavior in which the minimum gap occurs at a point; and (iii) band selective spectral characteristics, in which one band shows ``BCS" while the other shows ``BEC" behavior.
At large interactions, we find a Mott phase of rung bosons in which the filling is one boson for every two sites, half the typical density constraint for Mott insulators.
\end{abstract}

\maketitle

In many materials, superconductivity is seen to emerge directly from a gapped insulating state by tuning a parameter like magnetic field, disorder strength~\cite{havilandOnsetSuperconductivityTwodimensional1989} or doping~\cite{bollingerSuperconductorInsulatorTransition2011}.
This raises the question of how superconductivity can be born from an insulating state that has no Fermi surface~\cite{dobrosavljevicConductorInsulatorQuantum2012a,ghosalprb2000,ghosal-prb-2001,ghosal-prl-1998,bouadimSingleTwoparticleEnergy2011,swanson-prx-2014}? 
This question forces us to look beyond the standard
paradigm of Bardeen, Cooper and Schrieffer (BCS), in which superconductivity arises as a Fermi surface instability in a metal in the presence of effective attractive interactions. In the BCS regime, a gap opens up in the single particle density of states, though there is no gap for inserting pairs into the condensate. Further, the momentum resolved single particle spectral function shows that the locus of the minimum gap in momentum space is a closed contour along the underlying Fermi surface. How are these features changed as the insulator is approached?

\begin{figure}[htb!]
\centering
\includegraphics[width=0.5\textwidth]{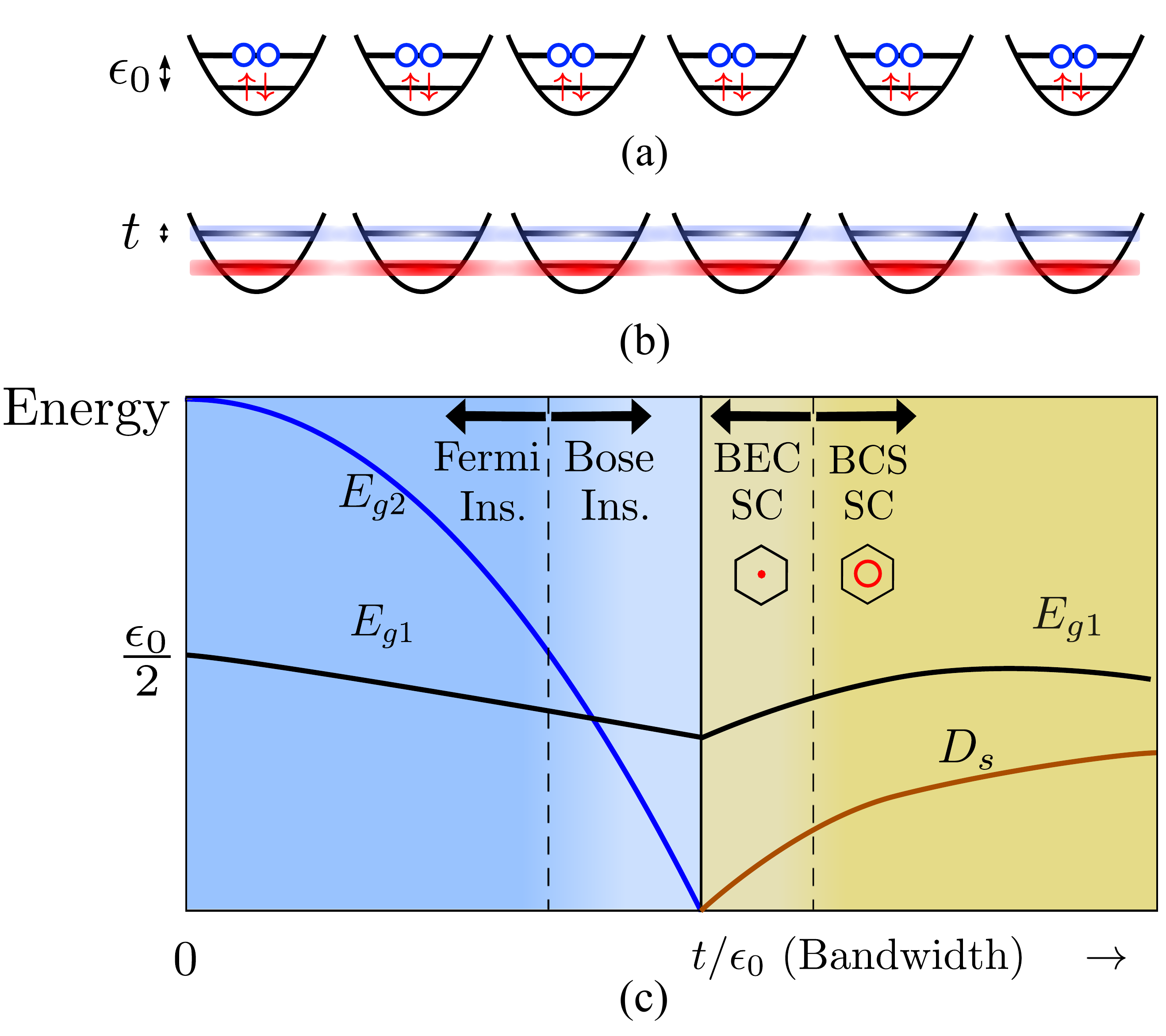}
\caption{Insulator to superconductor transition (SIT) in a two-band model: Atomic insulator with 2 orbitals per site separated by an energy $\epsilon_0$ [panel (a)] evolve into two non-overlapping bands upon including tunneling $t<<\epsilon_0$ forming a Fermi band insulator [panel (b)]. Upon including local attractive interactions $U$, the evolution of the energy scales as a function of $t/\epsilon_0$ at half filling is schematically illustrated in panel (c): The energy to add a single fermion $E_{g1}$ remains finite, the energy to add two fermions $E_{g2}$ starts out at twice $E_{g1}$ for $t=0$, but decreases and becomes smaller than $E_{g1}$ with increasing tunneling, crossing over from the Fermi band insulator to a Bose insulator (BI). The vanishing of $E_{g2}$ from the BI side indicates a SIT from the BI into a superconductor SC. In the SC, the superfluid density $D_s$, which is an energy scale in 2D, is finite and vanishes at the transition. Within the SC phase there is a crossover from the BEC regime where the locus of the minimum fermionic gap is a point to the BCS regime where the minimum gap locus is a contour in the Brillouin zone.
}
\label{f:SchematicEvol}
\end{figure}

We investigate this question in a simple disorder free two-band model with attractive interactions between fermions. The constraints on the model are that it must describe a band insulator and a metal to begin, in which superconductivity is the only symmetry-breaking phase encountered as interactions are turned on. Schematically our results are illustrated in Fig.~\ref{f:SchematicEvol}. 

We consider a lattice with two orbitals per site with two fermions occupying the lower orbital. As tunneling between the sites is introduced the orbitals broaden into bands forming the Fermi band insulator. With increasing tunneling, the bands overlap and the insulator transitions to a metal. 

In the presence of attractive interactions, a two-particle bound state and a two-hole bound state form within the insulating band gap.  Typically, the energy to create two excitations $E_{g2}$ costs twice the energy of creating a single excitation $E_{g1}$, however,  in the presence of attractive interactions there is a reduction due to the binding energy. With increasing tunneling, there is a crossover 
to a Bose Insulator defined by $E_{g2}<E_{g1}$ -- a regime where it is cheaper to create pairs rather than single particle excitations. The critical hopping strength where $E_{g2}=0$ marks the transition from the Bose insulator to a superconductor (SC) in which the superfluid stiffness $D_s$ starts to build up from zero at the transition. 

In this paper we identify a precise criterion to delineate the two regimes, BCS and BEC, even though it is a crossover with no change in symmetry. We show that there is a change in topology of the minimum-gap locus~\cite{randeriaCrossoverBardeenCooperSchriefferBoseEinstein2014}, which is readily observed in angle-resolved photoemission spectroscopy (ARPES) experiments. In the region close to the insulator-SC transition, the SC is in a BEC state with a minimum gap locus at $k=0$ implying that all the states in the Fermi volume have been affected by pairing. With increasing tunneling, the system crosses over to a BCS regime with a change in topology of the minimum gap locus that now lies along a closed contour at $\vec k_F$.

One of our remarkable 
observations is that of a band selective BEC-BCS crossover: we find an intermediate regime where 
the minimum gap locus becomes a finite momentum contour for the upper band, signaling a crossover from BEC to BCS regime, while it remains point-like on the other band. 
Generically in a two-band non-bipartite system, the crossover does not occur on both bands simultaneously: there is an intermediate regime where the band with higher (non-interacting) density of states at the gap edge is in the BCS regime, while the other can still be in the BEC regime. With further increase of tunneling, both bands evolve to have a minimum gap contour at finite momenta.
A second important observation is that in the Bose insulator close to the SIT, the single particle spectral functions are indistinguishable from those in the BEC-SC showing clear evidence of particle-hole mixing. These results are corroborated by detailed quantum Monte Carlo (QMC) simulations and analytical insights.

\section{Model and Outline of main results}\label{model}

\begin{figure}
   \centering
   \includegraphics[width=0.5\textwidth]{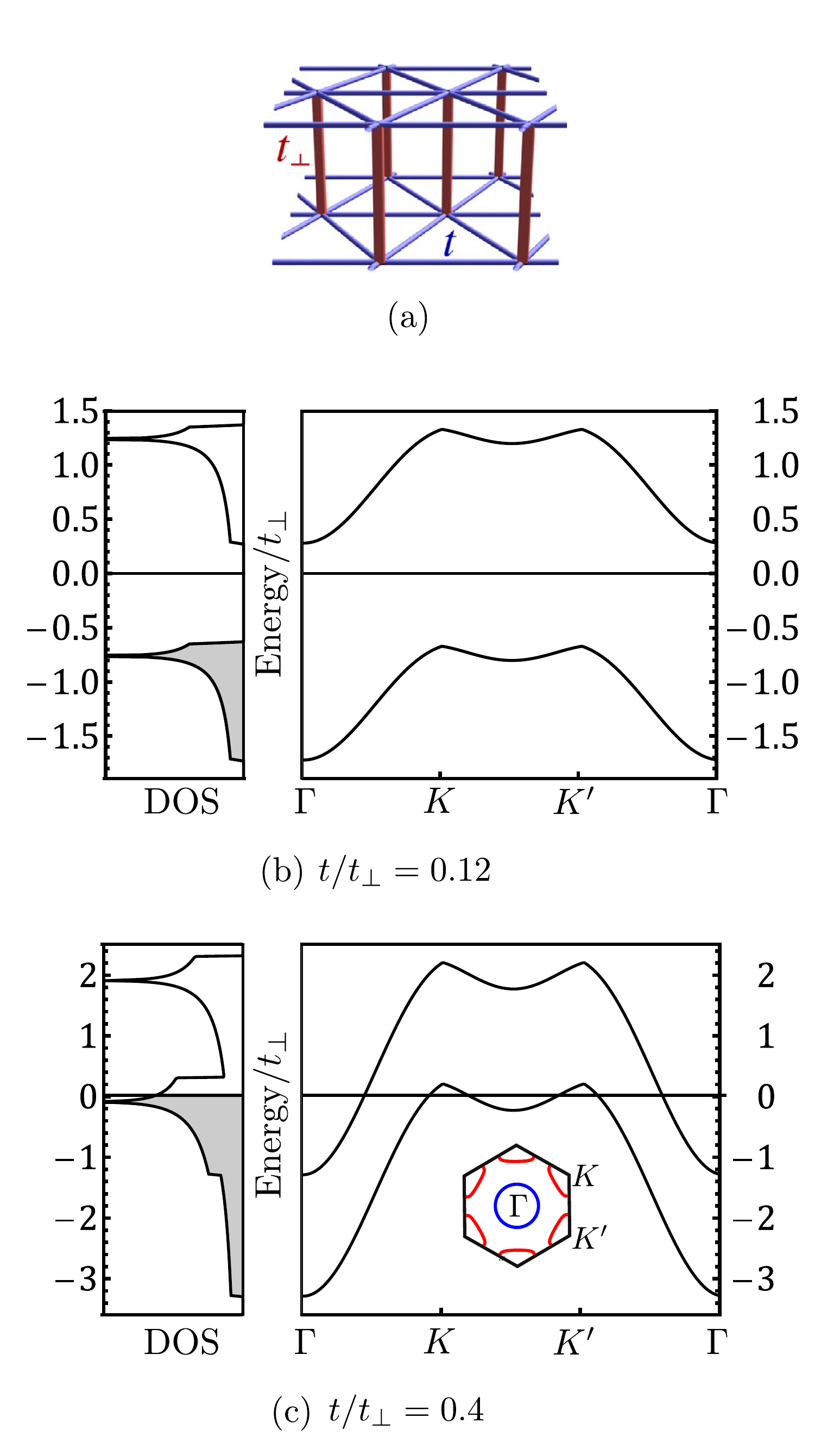}
   \caption{Insulator to metal transition at $U=0$ and half-filling. (a) Triangular lattice bilayer with nearest neighbour in-plane hopping ($t$) and nearest neighbour inter-layer hopping ($t_\perp$). (Figure from Ref.~\onlinecite{lohSuperconductorInsulatorTransitionFermiBose2016}.) (b,c) Density of states (left) and band dispersion (right) at (b) Insulator for $t=0.12t_\perp$; (c) Metal with electron and hole Fermi surfaces (shown in inset in blue and red respectively) for $t=0.4t_\perp$.}
   \label{f:DOSBands}
\end{figure}


On a bipartite lattice at half-filling, the attractive Hubbard model has SU(2) symmetry that results in a degeneracy between the superconducting ground state and the checkerboard charge density wave~\cite{millerCoexistenceSuperconductivityChargeDensity1993}. As a result, $T_c=0$. We therefore investigate the attractive Hubbard model on a non-bipartite lattice -- a triangular lattice bilayer that provides the minimal two bands needed to describe a band insulator (Fig. \ref{f:DOSBands}). This is described by the Hamiltonian $H=H_{KE} + H_U$ defined by:
\begin{align}
H_{KE} &= -t \sum_{\langle ij \rangle_\parallel,\sigma} \left( c_{i\sigma}^\dagger c_{j\sigma} + h.c. \right)
        \no \\ &\qquad
         -t_\perp \sum_{\langle ij \rangle_\perp,\sigma} \left( c_{i\sigma}^\dagger c_{j\sigma} + h.c. \right) - \mu \sum_{i\sigma} n_{i\sigma} \no\\
H_U &= -|U| \sum_i \left( n_{i\uparrow} - \frac{1}{2} \right) \left( n_{i\downarrow} - \frac{1}{2} \right)\label{fermHam}
\end{align}
where $c_{i\sigma}^\dagger$ creates a fermion at site $i$ with spin $\sigma$, which hops to the in-plane nearest-neighbour (NN) with amplitude $t$ or hops across the rung to the opposite layer with amplitude $t_\perp$, and $n_{i\sigma} = c_{i\sigma}^\dagger c_{i\sigma}$. The chemical potential is adjusted to maintain half-filling $n=N^{-1}\sum_{i\sigma} \langle n_{i\sigma}\rangle=1$, where $N$ is the number of sites.

$H_{KE}$ can be readily diagonalised in momentum space to obtain the dispersion $\epsilon_\veck = -2t\left(\cos(k_x) + 2 \cos(k_x/2) \cos(\sqrt{3}k_y/2)\right) - t_\perp \cos(k_z) - \mu$ where the two bands are labelled by $k_z = 0,\pi$. Fig.~\ref{f:DOSBands}~(c) shows the case of small $t/t_\perp$ in the absence of interactions. In this regie, the system is a band insulator with a bandgap $E_g= 2t_\perp-9t$. For $t>2/9 t_\perp$ (Fig.~\ref{f:DOSBands}~(d)), the bands overlap to form a compensated semimetal with an electron and a hole Fermi surface around the $\Gamma$ and $K$ points in the Brillouin zone respectively. These Fermi surfaces are unstable to pairing and result in a rich evolution of the spectral function that we discuss below. 

\begin{figure}
  \centering
  \includegraphics[width=0.45\textwidth]{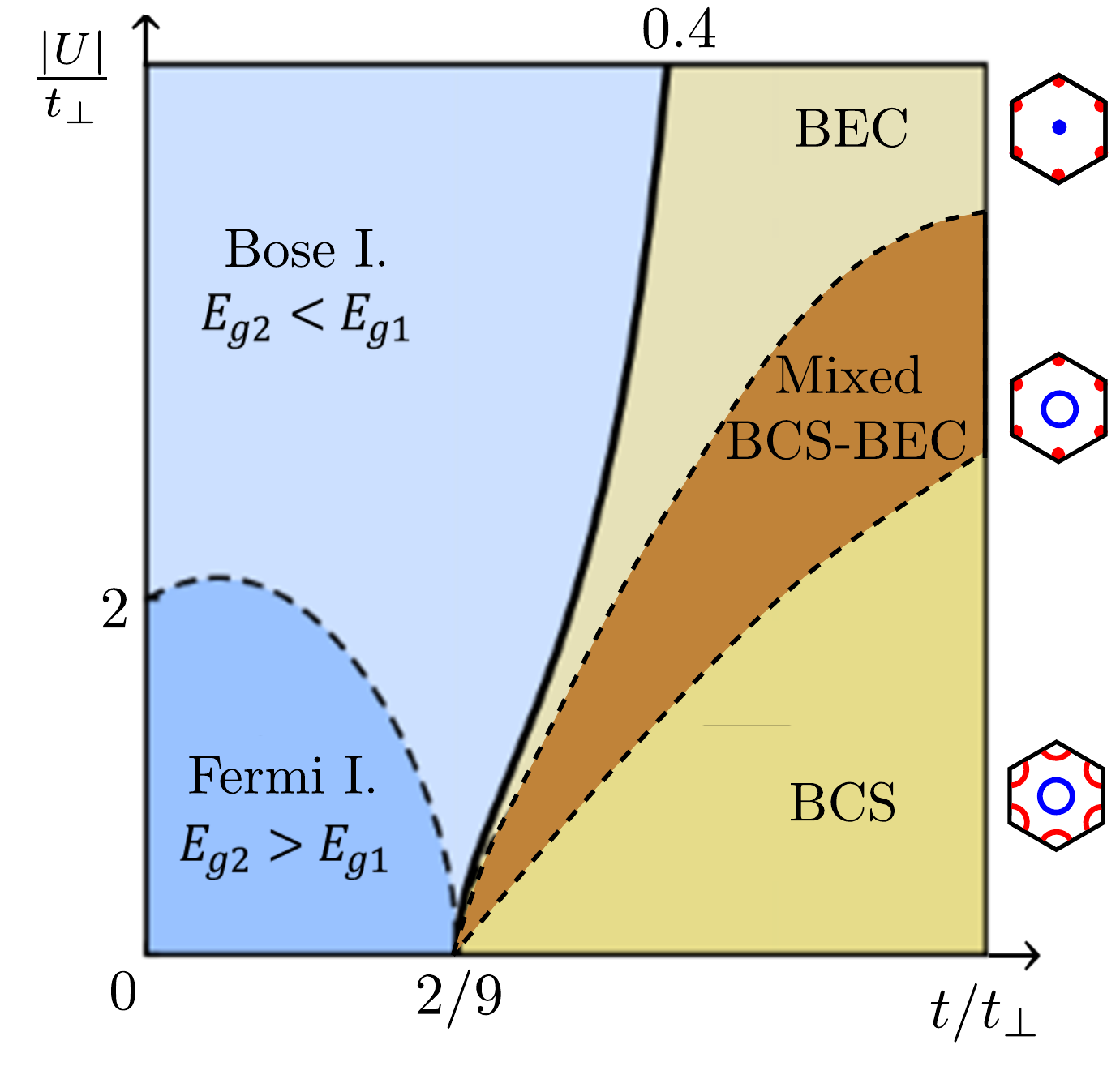}
  \caption{
  Phase diagram at half-filling and $T=0$ for the attractive Hubbard model on a triangular lattice bilayer: The blue phase is insulating, yellow indicates  a superconductor. The crossover from Fermi Insulator to Bose Insulator is defined by whether the lowest energy excitations are single fermions (charge $e$) or pairs (charge $2e$). The crossover from BEC superconductor to BCS superconductor is defined by whether the locus of minimum fermionic gap is a point or a contour in the Brillouin zone. The brown region is in a mixed BCS-BEC regime, where the min-gap locus is a point for one band and a contour for the other band. 
  At $U=0$, the transition from band insulator to metal occurs at $t/t_\perp=2/9$. Exact diagonalization in the limit of decoupled rungs fixes the crossover from Fermi to Bose insulator at $|U|/t_\perp=2$ in the $t=0$ limit. In the limit of strong coupling $|U|/t\gg1$, we obtain an estimate of the critical $t/t_\perp \approx 0.4$ from bosonic mean-field theory. 
  }
  \label{figPhaseDiag}
\end{figure}

\begin{figure}
  \centering
  \includegraphics[width=0.5\textwidth]{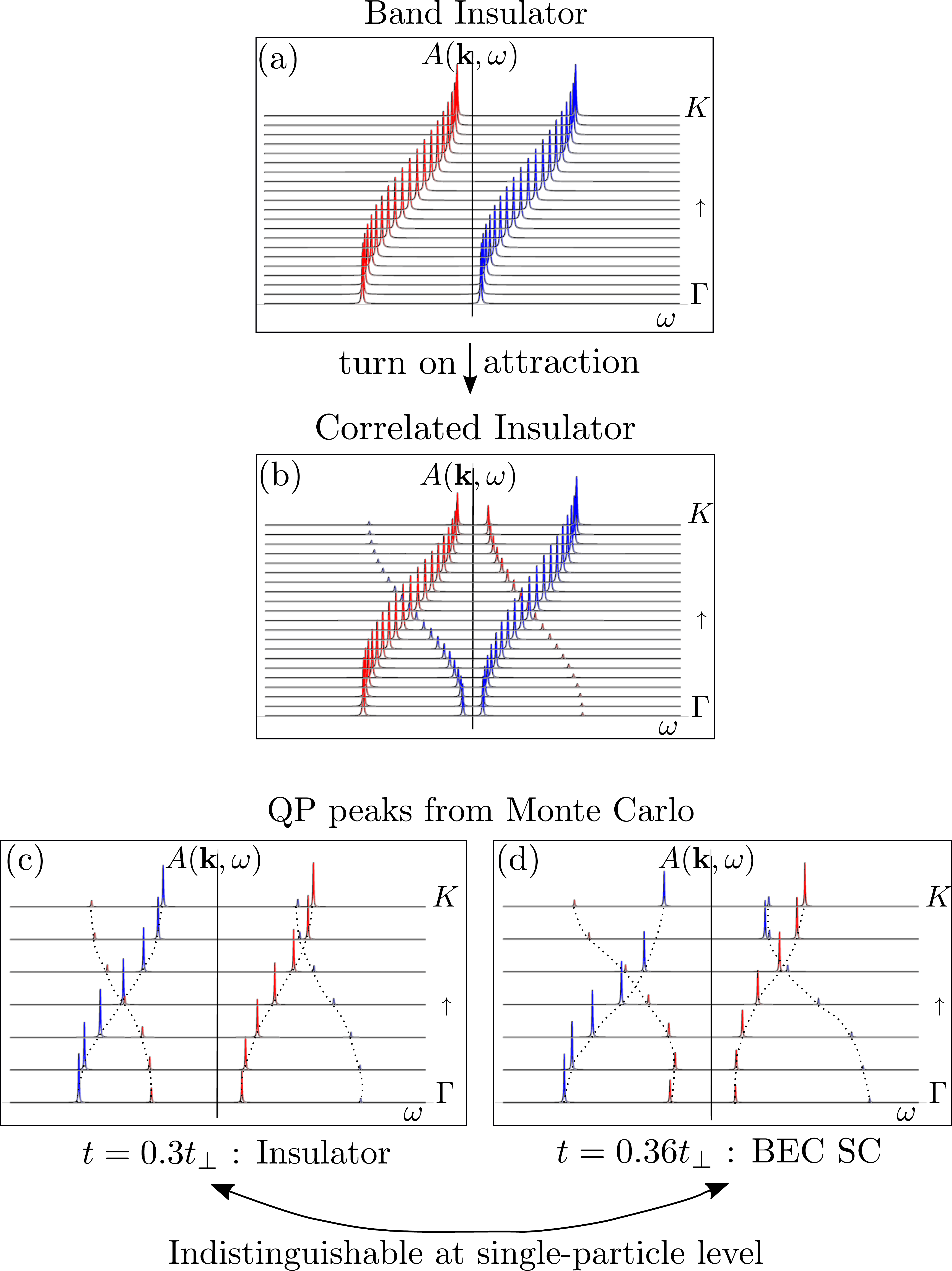}
   \caption{Doubling of quasiparticle poles in the incipient superconductor. Schematic plot of the spectral function $A(\mathbf{k},\omega)$ in (a) a band insulator with one pole in the spectral function for each band, and in (b) a correlated insulator where each band contributes two peaks to $A(\mathbf{k},\omega)$, due to the interactions mixing band eigenstates. (c) Estimates of quasiparticle energies and weights from QMC data in (c) the insulating regime at $t=0.3t_\perp$ and (d) the BEC superconductor at $t=0.36t_\perp$. Note that there is no qualitative change in the single-particle spectral function at the SIT. MC data at $|U|=4t_\perp, \beta t_\perp=12,\Delta\tau=0.05$ on a $12\times 12$ bilayer. Here, red(blue) identifies the band $k_z=0(\pi)$. }\label{f:Akwfig2}
\end{figure}

The model described above was introduced by Loh \emph{et.~al.}~\cite{lohSuperconductorInsulatorTransitionFermiBose2016} as a minimal model for studying the SIT in the absence of competing orders. 
They used BCS MFT and diagrammatics to obtain qualitative insights at weak coupling, and also then showed numerically using DQMC simulations at intermediate coupling strength that the gap in the single particle spectral function remained open through the SIT. When $t/t_\perp = 0$, exact diagonalization of the resulting two-site problem shows a crossover from Fermi insulator to Bose insulator at $|U|=2t_\perp$ (see Fig.~\ref{figPhaseDiag}). However, contrary to MFT which predicts a superconductor-insulator transition in this atomic limit, the exactly computed gap to pair excitations does not close as interaction is increased, indicating that the atomic insulator does not become a superconductor. 

In this paper, we present exact results for the single-particle and two-particle Green's function in the atomic limit where $t=0$. We clearly see how the spectral function of the correlated insulator at finite $|U|$ interpolates between the spectral function of the band insulator and the familiar BCS form of a superconductor. This exact expression illuminates the smooth evolution of the spectral function that we observe in our QMC simulations. 

At intermediate coupling $|U|/t_\perp=4$, we undertake a detailed study of the SIT and the BCS-BEC crossover using QMC simulations. Using the imaginary-time dependence of the corresponding Green's functions, without analytic continuation, we clearly show that the single-particle gap remains open while the two-particle gap closes at the transition. We resolve for the first time, a multi-band BCS-BEC crossover identified by the classic signature in the topology of the min-gap locus. We identify a ``band selective BCS-BEC" regime in which one-band is in a BEC regime and the other is in a BCS regime, as indicated in the phase diagram in Fig.~\ref{figPhaseDiag}. We also identify clear signatures of a pairing pseudogap regime above $T_c$ in the superconductor.

The insights gleaned from this model are quite general. In Fig.~\ref{f:Akwfig2} we discuss the evolution of the quasiparticle spectrum across the phase diagram. This is captured by the spectral function $A(\mathbf{k},\omega)$ which indicates the probability of finding a single-particle excitation with energy $\omega$ and momentum $\mathbf{k}$.


Fig.~\ref{f:Akwfig2}(a) shows the spectral function of the non-interacting band insulator with one pole per band whose energy tracks the band dispersion $\epsilon_\mathbf{k}$. As we turn on interactions, the spectral function for each band develops two poles - one each at positive and negative energy, as in Fig.~\ref{f:Akwfig2}(b). This reflects the partial occupation of the momentum eigenstates at zero temperature in the presence of interactions. These schematic figures are supported by exact analytical expressions derived in Sec.~\ref{atomic} for the spectral function in the atomic limit ($t=0$) as a function of interaction strength. In Fig.~\ref{f:Akwfig2}(c), we show QMC estimates of the quasiparticle peaks and their weights for the insulator at $|U|=4t_\perp$. Note that this spectral function {\it in the insulating regime} is remarkably qualitatively similar to the superconducting spectral function in the BEC regime, shown in Fig.~\ref{f:Akwfig2}(d). {\it The mixing of the particle and hole spectral weights precedes the SIT}. As we further increase $t/t_\perp$, the spectral function smoothly evolves into the familiar BCS form with the back-bending of the quasiparticle peaks. 

The multi-band BCS-BEC crossover has been studied in a variety of settings~\cite{zhaoBCSBECCrossoverTwoDimensional2006,mondainiMottinsulatorSuperconductorTransition2015,chubukov_superconductivity_2016,innocentiResonantCrossoverPhenomena2010,iskinTwobandSuperfluidityBCS2006,salasnichScreeningPairFluctuations2019,tajimaEnhancedCriticalTemperature2019} and is discussed as a promising route to higher $T_c$~\cite{salasnichScreeningPairFluctuations2019} via a deep band in a BCS regime suppressing phase fluctuations and a shallow band in a crossover regime providing a high density-of-states as suggested in Ref.~\onlinecite{hazraBoundsSuperconductingTransition2019}. We provide the first quantum Monte Carlo evidence of such a band-selective crossover in Fig.~\ref{f:Gapstory}.

We also study the strong-coupling limit of this model using an effective boson Hamiltonian. Using MFT we identify a Mott-insulator to superfluid transition at $t/t_\perp=0.4$, at \textit{half}-filling. 

\section{Determinantal Quantum Monte Carlo}\label{DQMC}

In this section, we identify the essential features of the SIT and the BCS-BEC crossover at $|U|/t_\perp=4$ and half-filling, using sign-problem free Determinantal Quantum Monte Carlo (DQMC) simulations~\cite{blankenbeclerMonteCarloCalculations1981}.

In DQMC, the on-site Hubbard interaction between fermions is replaced by a coupling to an auxilliary Hubbard-Stratonovich field, which varies in space and imaginary time. The fermions are then integrated out to obtain an action for the Hubbard-Stratonovich field configurations, which are then sampled according to the Metropolis algorithm. The problem of interacting fermions in two dimensions is thus mapped onto a $(2+1)D$ problem of classical fields.
This allows unbiased, statistically exact, sign-problem free calculation of observables in the interacting fermion problem on finite sized lattices at finite temperature.

\subsection{Single Particle Gaps}\label{1gap}

\begin{figure*}
\centering
\includegraphics[width=0.85\textwidth]{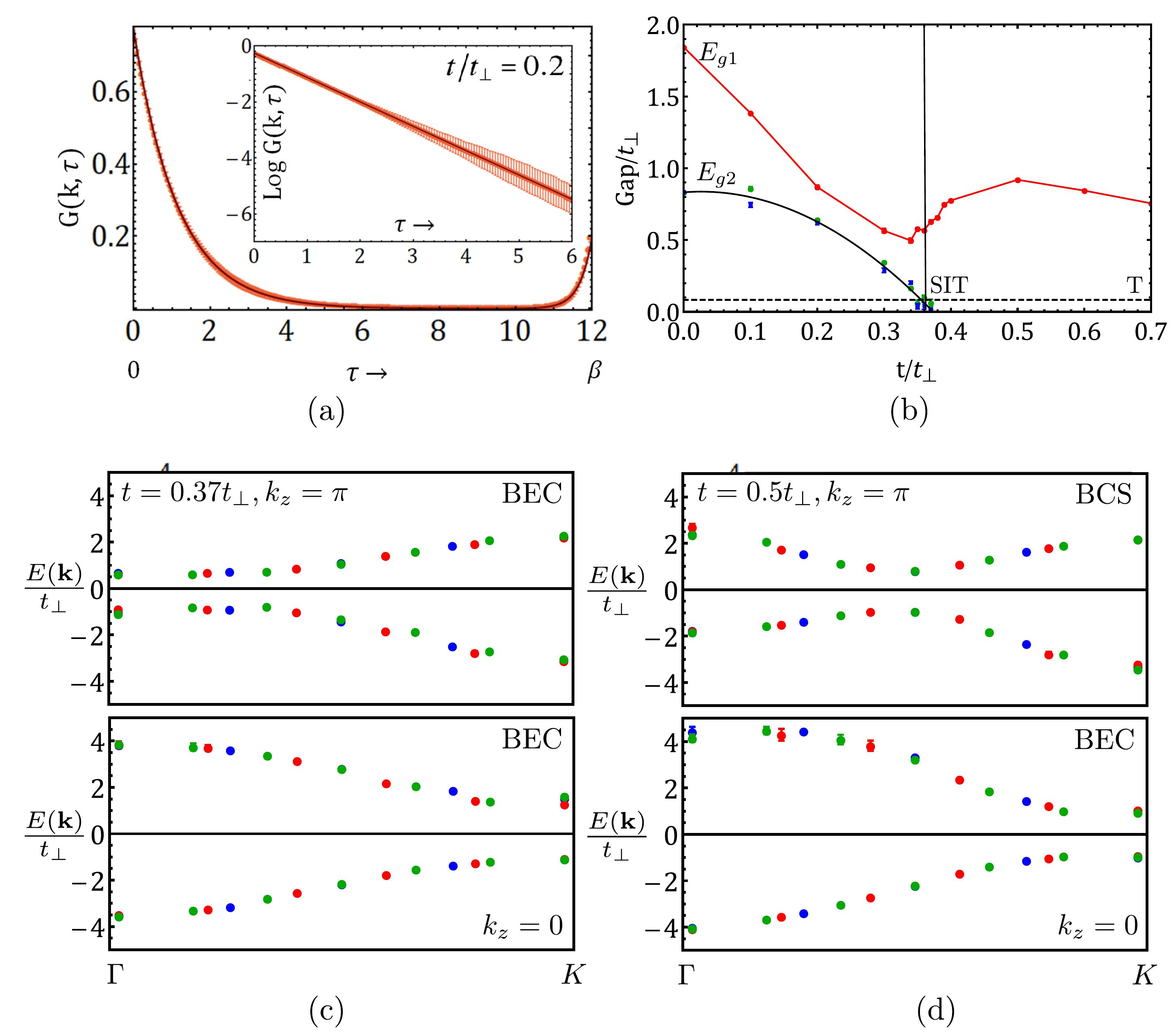}
\caption{Single particle and two particle gaps from Monte Carlo simulations:
(a) Gap extraction from single particle Green's function $G(\veck=(0,0,\pi),\tau)$. The black curve is a fit to the DQMC data in red from a $10\times10$ bilayer. The slope of $\log G(\veck,\tau)$ (shown in inset) near $\tau=0(\beta)$ gives the gap to particle (hole) excitations. See text for details. (b) Energy gaps across the SIT. The single particle gap $E_{g1}$ remains finite, while the two-particle gap $E_{g2}$ goes soft at the transition. The non-monotonic variation of $E_{g1}$ is explained in the text. $E_{g1}$: $10\times10$ bilayer (red), $E_{g2}$: $8\times8$ (blue) and $6\times6$ (green) bilayers. The black curve is intended as a guide to the eye. (c,d) BCS-BEC crossover  on $k_z=\pi$ band (upper panels) while the $k_z=0$ band (lower panels) remains in a BEC regime: Single particle gaps from DQMC on the $k_z=\pi$ (upper) band along the cut through the Brillouin zone shown in blue in the insets for $8\times8$ (blue), $10\times10$ (red), $12\times12$ (green) bilayers. 
DQMC data is at half-filling with $|U|=4t_\perp, \beta t_\perp=12, \Delta\tau=0.05$. 
}
\label{f:Gapstory}
\end{figure*}

We estimate the single particle gap at different momenta from the imaginary time dependence of the single particle Green's function $G(\mathbf{k},\tau)=\langle c_{\veck\sigma}(\tau) c^\dagger_{\veck\sigma}(0) \rangle$. This is related to the spectral function $A(\veck,\omega)$ by
\begin{align}
G(\veck,\tau)=\int_{-\infty}^\infty d\omega	\frac{e^{-\omega \tau}}{1+e^{-\beta\omega}} A(\veck,\omega).
\end{align}
By fitting the Monte Carlo data for $G(k,\tau)$ as in Fig. \ref{f:Gapstory}(a), we estimate the quasiparticle energies and the spectral weights in each quasiparticle pole (details in Appendix~\ref{a:gapextraction}). The momentum resolution of this technique is discussed below, first we focus on some global features.

We define $E_{g1}$ as the smallest gap over the Brillouin zone in the two bands and find that this gap remains finite across the transition (Fig. \ref{f:Gapstory}(b)). 
By tracking the 
single-particle gap across the superconducting transition, we are able to conclusively rule out the presence of any intervening metallic phase where the single-particle gap also closes. 
At the single particle level, there are no low energy charge degrees of freedom in the superconductor or the insulator. 
The non-monotonic dependence of $E_{g1}$ with increasing $t/t_\perp$ can be intuitively understood using insights from mean-field theory at small $|U|/t_\perp$ (c.f. Fig. 2 of Ref.~\onlinecite{lohSuperconductorInsulatorTransitionFermiBose2016}). 
The initial decrease with increasing bandwidth in the insulator is understood as a linearly decreasing band-gap. 
In the BEC regime of the superconductor just past the transition, the non-interacting band-gap $E_{g0}$ and the pairing order parameter $\Delta$ add in quadrature within MFT: $E_{g1}=\sqrt{(E_{g0})^2 + \Delta^2}$. 
This explains the initial increase of the one-particle gap after the transition. 
Eventually, at large $t/t_\perp$, the order parameter decreases as the non-interacting density of states decreases with increasing bandwidth. 

The momentum resolution of DQMC allows us to identify the locus of points in the Brillouin zone where the single-particle gap is minimum. 
Within weak-coupling BCS theory, this mimimum gap locus is a contour that coincides with the non-interacting Fermi surface, which produces the well-known coherence peak in the density of states at the gap-edge. 
We identify the crossover to a BEC regime by the criterion that the minimum gap locus on a band shrinks to a point. 
The density of states in this regime has a jump discontinuity in two dimensions that is inherited from the jump in the non-interacting density of states at the band-edge in two dimensions. 
This qualitative distinction between the gap-edge singularity across the BCS-BEC crossover has only recently been pointed out in the literature~\cite{lohSuperconductorInsulatorTransitionFermiBose2016,chubukov_superconductivity_2016}. 

In Fig. \ref{f:Gapstory}(c,d), we plot the single-particle gap on the $k_z=\pi$ band along the $\Gamma-K$ cut through the Brillouin zone on different lattice sizes (shown in different colours). Near the SIT (Fig. \ref{f:Gapstory}(c)), the gap minimum is at the $\Gamma$ point and the superconductor is in the BEC regime. As we increase $t/t_\perp$, the minimum gap shifts to a finite $\veck$ and the locus of minimum single-particle gap is indicated on the Brillouin zone in red (Fig. \ref{f:Gapstory}(d)). We thus observe, for the first time in Monte Carlo simulations, a BCS-BEC crossover in a multi-band system, resolved by the min-gap locus evolving from a point to a contour. This qualitative change in the topology of the minimum gap locus has recently been observed in ARPES experiments on Fe$_{1+y}$Se$_x$Te$_{1-x}$~\cite{rinottTuningBCSBECCrossover2017a}. 

The minimum gap locus on the $k_z=0$ band remains at the $K$ and $K'$ points and does not evolve into a contour for the parameter range shown in Fig. \ref{f:Gapstory}(c,d) for the lattice sizes we have considered. 
Intuition from MFT indicates that this band-selectivity of the crosover can be understood in terms of the low-energy density of states of the underlying bandstructure. 
On non-bipartite lattices, the density of states is different at the two band edges on either side of the gap. For instance, 
Fig.~\ref{f:DOSBands}(c) shows that in our model, the density of states at the gap edge is smaller in the lower band. As a result, numerically solving the number equation for half-filling in the superconducting state, $\sum_{\alpha}\int d\epsilon {\, } N_{\alpha}(\epsilon) \frac{\epsilon-\mu}{E}\tanh \frac{\beta E}{2}=0$, with 
$E = \sqrt{(\epsilon-\mu)^2+\Delta^2}$
shows that the chemical potential $\mu$ is closer to the upper band. Here $N_{\alpha}(\epsilon)$ is the density of states in band $\alpha$ at energy $\epsilon$; $\beta$ is the inverse temperature. If we take the limit of zero temperature and vanishing \textit{bandgap}, at a given value of pair potential and bandwidth, by keeping $t,\Delta$ fixed and reducing $t_\perp$, we obtain the number equation:
\begin{align}
    \left[ N_{k_z=0}\int_{-\infty}^{\epsilon_{K,k_z=0}} + N_{k_z=\pi}\int^{\infty}_{\epsilon_{\Gamma,k_z=\pi}} \right] d\epsilon \frac{\epsilon-\mu}{E} \approx 0
\end{align}
where we have approximated the ($t_\perp$-independent) density of states of each band by its average near the band edge. Clearly, this equation is satisfied when the chemical potential leans toward the band with the lower density of states at the band edge. In the limit of zero bandgap, this band therefore has an underlying Fermi surface and is in the BCS regime, while the min-gap locus on the other band is a point, characteristic of the BEC regime.

In Appendix \ref{Akw}, we explore some alternative metrics to delineate the crossover and discuss their merits and demerits.

\subsection{Two particle gap}\label{2gap}

The absence of low lying fermionic modes precludes the usual BCS understanding of superconductivity as an instability of the Fermi surface. In absence of a Fermi surface, it is natural to ask what precipitates this quantum phase transition in the insulator. Following earlier work on attractive Hubbard models~\cite{lohSuperconductorInsulatorTransitionFermiBose2016,tsuchiyaHiggsModeSuperfluid2013} and on the disorder-driven SIT~\citep{bouadimSingleTwoparticleEnergy2011} we conclusively show that it is the gap to two-particle (charge $2e$) excitations that goes soft at the SIT and leads to superconductivity.

We extract the two-particle gap from the two-particle Green's function $P(\mathbf{q},\tau)=N^{-2} \sum_{i,j}\langle  c_{j\downarrow} c_{j\uparrow}(\tau) c^\dagger_{i\uparrow} c^\dagger_{i\downarrow} (0) \rangle e^{-i\mathbf{q}.(\mathbf{r}_i - \mathbf{r}_j)}$ where $N$ is the number of sites. The $\mathbf{q}=0$ component of this object is the propagator for zero center-of-mass momentum pairs
\begin{align}
P(\mathbf{q}=0,\tau)=\frac{1}{N^{2}}\sum_{k,k^{'}} \langle c_{-k\downarrow}^\phdag c_{k\uparrow}^\phdag (\tau) c_{k^{'}\uparrow}^\dagger c_{-k^{'}\downarrow}^\dagger (0) \rangle.
\end{align}
The energy scale of $P(\mathbf{q}=0,\tau)$ therefore corresponds to the minimum energy cost of introducing a fermion pair into the system. We extract this energy scale by fitting the DMQC data for $P(\mathbf{q}=0,\tau)$; 
(details in Appendix~\ref{a:gapextraction}).

The smaller of the particle and hole gaps thus extracted is denoted as the two-particle gap $E_{g2}$ and is shown in Fig. \ref{f:Gapstory}(b). We find that it goes soft near the SIT. 

\subsection{Compressibility and Superfluid Density}\label{DirectSIT}

In the discussion below, we confirm that there is a direct transition from insulator to superconductor with no intervening metallic phase as we increase the bandwidth by tuning the ratio $t/t_\perp$. 

\begin{figure*}
   \centering
   \subfigure[\label{f:Tcdemo}]{
   \includegraphics[width=0.32\textwidth]{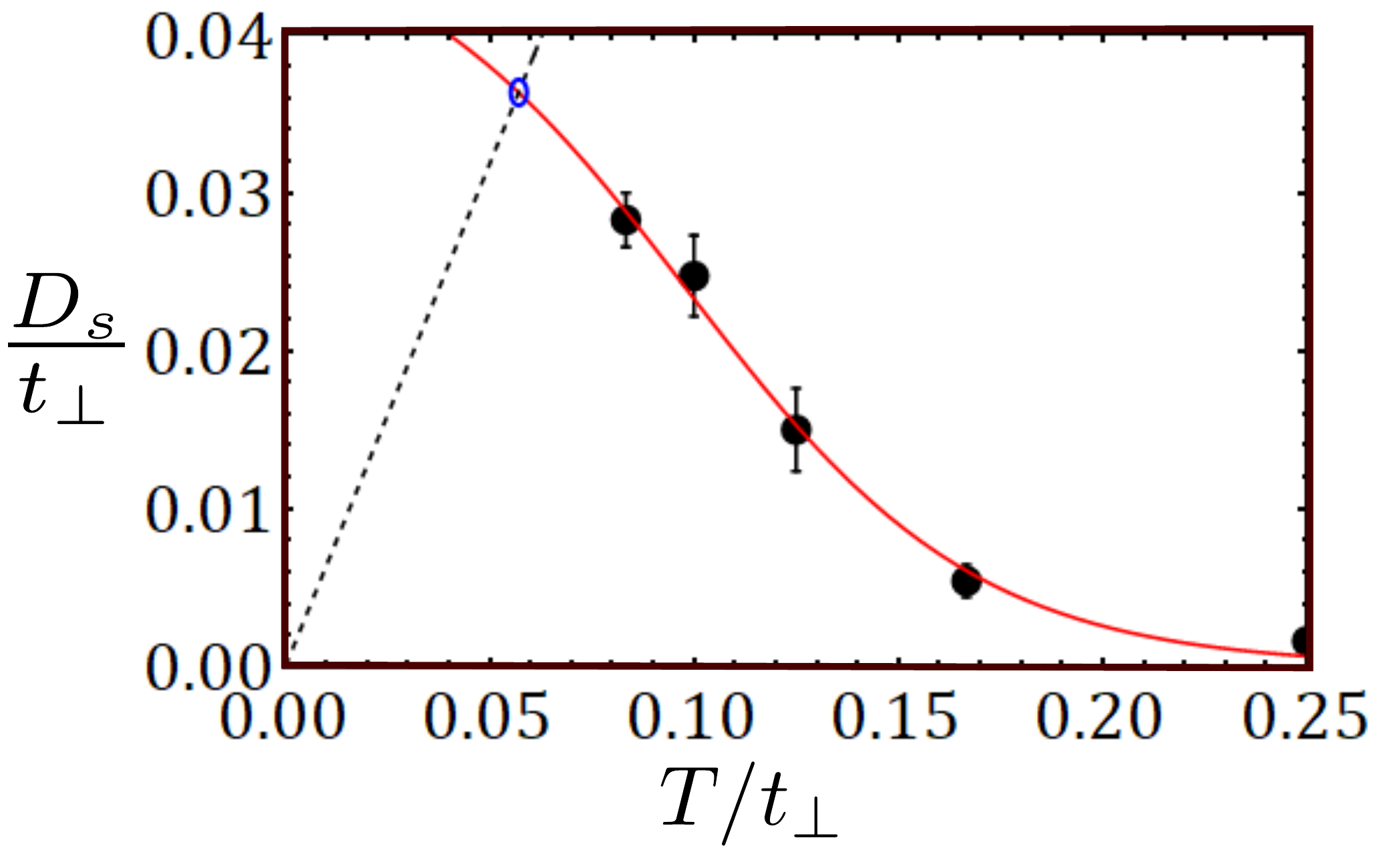}
   }
   \subfigure[\label{f:DirectSIT}]{
   \includegraphics[width=0.32\textwidth]{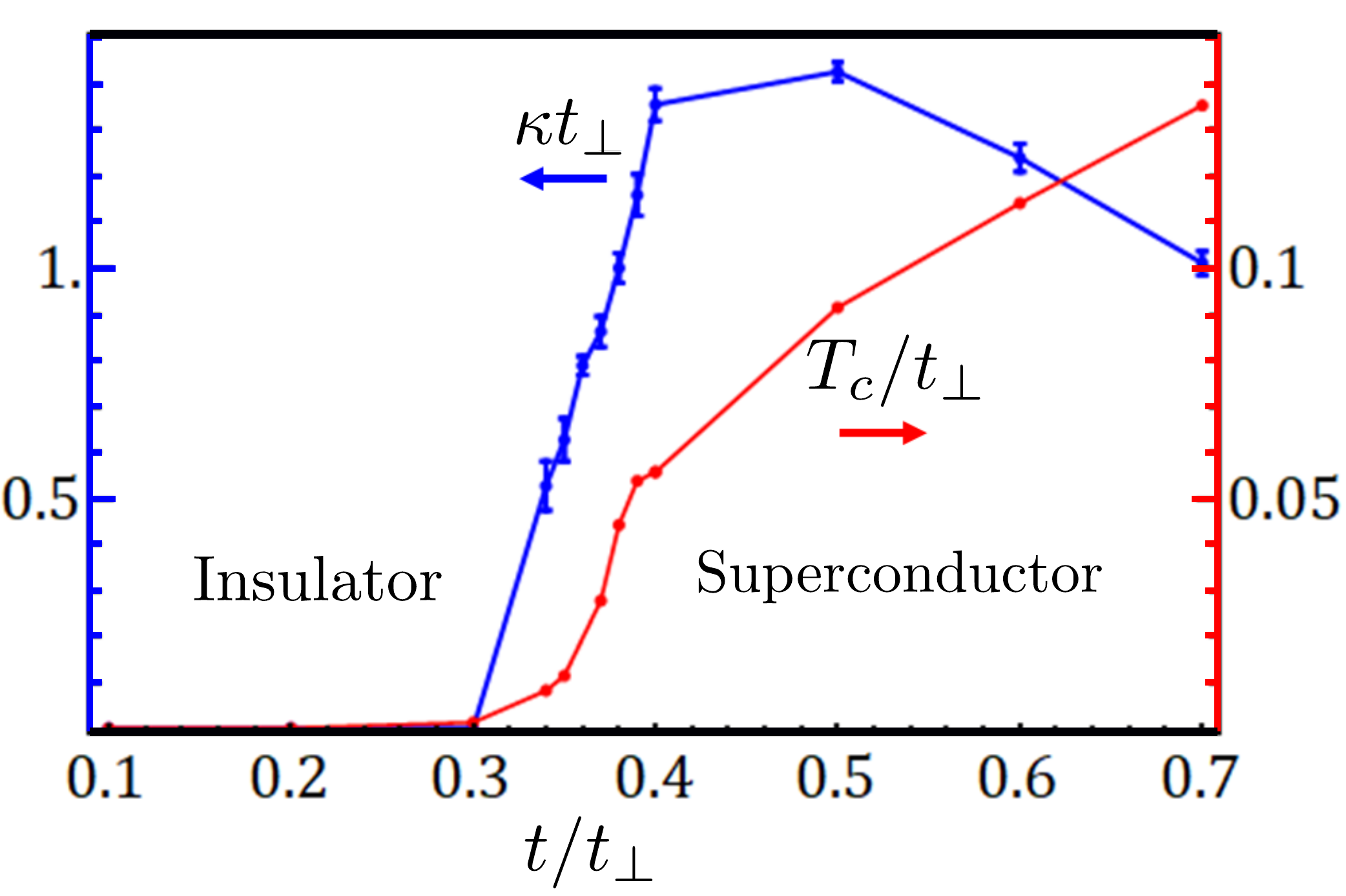}
   }
   \subfigure[\label{f:pseudogap}]{
   \includegraphics[width=0.3\textwidth]{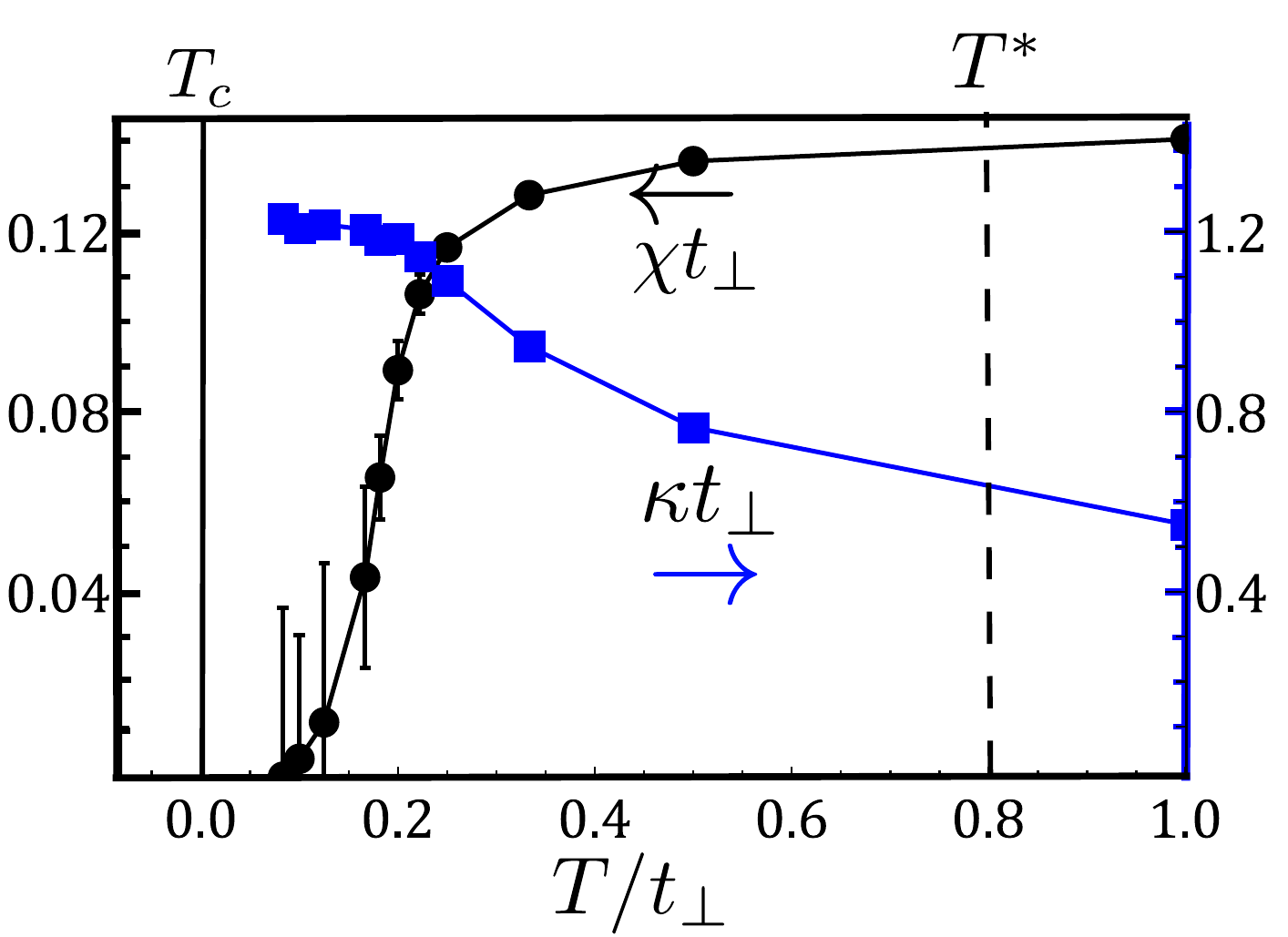}
   }
   \caption{
   \subref{f:Tcdemo} Estimation of $T_c$ from the Nelson-Kosterlitz jump condition. The red curve is an extrapolation of DQMC data (black) for $D_s$ at different temperatures $T$. The dashed line represents $2T/\pi$ and its intersection with the red curve (blue circle) is the estimated $T_c$. Data shown for $t=0.4t_\perp,\Delta \tau=0.05$ for a $10\times 10$ lattice. 
   \subref{f:DirectSIT} Direct transition from insulator to superconductor: The zero-temperature compressibility $\kappa$ (blue) is non-zero in the region where the superconducting $T_c$ (red) is finite. The former is obtained by extrapolating finite temperature compressibility data from DQMC to $T=0$ using a simple polynomial fit. The latter is obtained by similarly fitting finite temperature superfluid stiffness from DQMC to find where the Nelson-Kosterlitz jump condition is satisfied.
   \subref{f:pseudogap} Suppression of spin susceptibility (black) in the pairing pseudogap regime. Below the pair-breaking energy scale $T^*$, low energy fermions are paired up into spin-0 bosons, which results in suppression of the spin degree of freedom. This results in vanishing spin susceptibility, even as charge susceptibility (compressiblity in blue) increases with decreasing temperature.The data shown here is for $t/t_\perp=0.6$ on a $10\times 10$ bilayer.
   All three figures correspond to half-filling and $|U|=4t_\perp$.
   }
   \label{f:SIT}
\end{figure*}

An insulating state is characterized by vanishing compressibility $\kappa=n^{-2} (dn/d\mu)$, which we estimate by measuring the density $n$ for different chemical potential $\mu$ and obtaining $n(\mu)$. We obtain estimates at $\beta=12,10,8,6,4$ and extrapolate $\kappa(n=1,T)$ to zero temperature. We find (Fig. \ref{f:DirectSIT}) that the zero-temperature compressibility is finite beyond $t/t_\perp \sim 0.3$.

To establish that this compressible state is a superconductor, we calculate superfluid stiffness $D_s$ from the Kubo formula for the transverse limit of the current-current response $\chi$ to a static vector potential
\begin{align}
D_s=\widetilde{D}-\frac{\hbar^2}{4e^2}\chi_{j_xj_x} (q_x=0,q_y=\frac{2\pi}{L},\omega=0)
\end{align}
where $L$ is the linear size of the system, the diamagnetic response $\widetilde{D}$ is 
\begin{align}
\widetilde{D}=\frac{1}{4 L^2} \sum_{\veck \sigma} \partial_{k_x}^2 [\epsilon_\veck n_\veck]
\end{align} 
and the paramagnetic response is given by the current-current correlation function $\chi_{j_x j_x}$.	

We measure $D_s(T)$ at $\beta=12,10,8,6,4$ and estimate (as shown in Fig. \ref{f:Tcdemo}) the superconducting transition temperature $T_c$ from the Nelson-Kosterlitz jump condition $D_s(T_c^-) = 2 T_c/\pi$. We find (Fig. \ref{f:DirectSIT}) that the ground state is a superconductor beyond $t/t_\perp=0.3$ where the system ceases to be an insulator. (Note that the phase stiffness of the charge $2e$ bosons that enters the Nelson-Kosterlitz relation differs from the energy scale related to the superfluid weight defined by Scalapino, White and Zhang~\cite{scalapinoInsulatorMetalSuperconductor1993} by a factor of 4; see Appendix A of Ref.~\onlinecite{hazraBoundsSuperconductingTransition2019}.)

\subsection{Pairing Pseudogap}\label{pseudogap}

In the superconductor, the critical energy scale at zero temperature is the superfluid stiffness, which measures the energy cost of phase fluctuations. Away from the weak-coupling BCS regime in 2D, it is the superfluid stiffness that sets the scale for the superconducting critical temperature $T_c$. Since the single-particle gap remains finite through the transition while the superfluid stiffness vanishes, on the superconducting side of the transition we expect a pairing pseudogap regime at temperatures above $T_c$ but below the single-particle energy scale $T^*$, set by the minimum fermionic gap. In this regime, the coherence between pairs is destroyed by thermal phase fluctuations but single particle excitations are still gapped. We emphasize that this pairing pseudogap regime necessarily accompanies a continuous SIT where the superfluid stiffness vanishes smoothly at the transition.

In our model, in this regime, the fermions are paired up into spin-0 singlets. This results in a suppression of the spin susceptibility below the crossover scale $T^*$ shown in Fig. \ref{f:pseudogap} and a concomitant reduction of the low energy density of single particle states even in the normal state above $T_c$. Note that a necessary criterion for the pairing pseudogap phase is that single-particle excitations are gapped, which may not be valid~\cite{maletzUnusualBandRenormalization2014a} for the strongly coupled iron-chalcogenide superconductor FeSe where several works have reported a neglible pairing pseudogap regime above $T_c$~\cite{yangBCSlikeCriticalFluctuations2017,hanaguriQuantumVortexCore2019} 
despite the pairing gap $\Delta \sim \ef$ as expected in the BCS-BEC crossover regime.


\section{Exact diagonalization in the two-site limit}\label{atomic}

In the limit of vanishing in-plane hopping ($t$), the lattice decouples into a set of two-site Hubbard models on each vertical rung. This exactly-solvable limit allows us to address what happens to an atomic insulator as the interaction strength is continuously increased.   

The Hamiltonian in Eq.~\ref{fermHam} in this two-site limit takes the form,
\begin{align}
H=-t_\perp \sum_\sigma\left( c_{1\sigma}^\dagger c_{2\sigma} \right) - |U| \sum_{i=1,2} (n_{i\uparrow}-1/2)(n_{i\downarrow}-1/2)\label{atomicHam}
\end{align}
where $c_{i\sigma}^\dagger (c_{i\sigma})$ creates (destroys) an electron with spin $\sigma$ on the site $i=1,2$. Analytical calculations in this limit show~\cite{lohSuperconductorInsulatorTransitionFermiBose2016} that the ground state undergoes a smooth crossover from a band insulator at $U=0$ to a Mott insulator at $|U|\gg t_\perp$ (We will look closely at the latter in Sec.~\ref{largeU}).

We calculate the single particle spectral function and this reveals an interesting feature of the insulating state at finite $|U|$. The single particle spectral function is defined in terms of the retarded Green's function $A_{\alpha}(\omega)={-\rm Im } G^R_{\alpha}(\omega)/\pi$ where $\alpha=+(-)$ denotes the bonding (anti-bonding) orbital on the rung, corresponding to $k_z=0(\pi)$. At $T=0$ (details in Appendix~\ref{a:atomic}),
\begin{align}
A_{+}(\omega)= & \mathcal{N}_-^2 \delta(\omega - (t_\perp + E_0)) \no\\
					 &+ \mathcal{N}_+^2 \delta(\omega + (-t_\perp +E_0)) \\
A_{-}(\omega)= & \mathcal{N}_+^2 \delta(\omega - (-t_\perp + E_0)) \no\\
					 &+ \mathcal{N}_-^2 \delta(\omega + (t_\perp +E_0)) \label{atomicAkw}
\end{align}
where $\mathcal{N}_+= \cos(\pi/4- \theta/2)$, $\mathcal{N}_-= \sin(\pi/4- \theta/2)$, $\theta=\tan^{-1} (4t_\perp/|U|)$ and $E_0=\sqrt{U^2 + 16t_\perp^2}/2$. 
Unlike a band insulator which has one pole on each band corresponding to a particle or hole excitation, the $|U|>0$ insulator has a pole both at positive and negative energy. 
Unlike the BCS limit of a superconductor, the positive and negative energies are not equal. 
The reason for the apparent particle-hole mixing in Eq.~\ref{atomicAkw} is that both bands ($k_z=0,\pi$) are partially occupied in the ground state due to interband pair hopping processes at any finite $U$. As a result, there is finite probability of creating a particle or a hole in each band.

The exact spectral function of the correlated insulator computed in the atomic  limit bears qualitative resemblance to a superconducting spectral function and reinforces the intuition that the single particle excitations evolve smoothly through the SIT. Indeed, this smooth evolution of the single particle Green's function across the SIT is clearly seen in the Monte Carlo results at finite $t/t_\perp$ in Sec.~\ref{DQMC}. Beyond the SIT, the two poles evolve to equal and opposite energies, and the quasiparticle weights evolve smoothly to the well-known BCS coherence factors in the BCS limit.

In the limit of vanishing in-plane hopping, we find that the ground state for arbitrary $|U|/t_\perp$ generically has two fermions per rung. For a band insulator, the gap to two-fermion excitations is twice the gap to single-fermion excitations. In contrast, for $|U|>2t_\perp$, the lowest gap to single-fermion excitations $E_{g1}=E_0-t_\perp$ exceeds the gap to pair excitations $E_{g2}=E_0-|U|/2$ , beyond which point the system is defined to be a Bose Insulator. We note that the two-particle gap $E_{g2}$ does not close for any value of $|U|/t_\perp$. We also find that the single-particle spectral function on each band has two poles, as in a superconductor, which distinguishes the insulator from a simple band insulator. 

\section{Bosonic MFT in the strong coupling limit}\label{largeU}

Having established that the insulating state at small $t/t_\perp$ undergoes a crossover from a Fermi insulator to a Bose insulator with increasing $|U|/t_\perp$, we focus on the nature of the insulator at $|U| \gg t_\perp \gg t$. The low energy physics in this limit is governed by the dynamics of tightly bound fermion pairs (bosons) and at the filling of half a boson per site, one would not normally expect a Mott insulator. However, the on-site pairs delocalise over the vertical rungs and the ground state in this limit does turn out to be a Mott insulator of one boson per rung. The zero-temperature phase transition, in this limit, amounts to Bose condensation of these rung-bosons, and is qualitatively understood within a framework similar to the mean-field theory of point bosons developed by Fisher, Seshadri and others~\cite{fisherBosonLocalizationSuperfluidinsulator1989,sheshadriSuperfluidInsulatingPhases1993}.

In the limit of strong attractive interaction $|U|\gg t,t_\perp$, the low energy states are those in which all fermions are paired up into on-site bosons defined by the creation operators $b_i^\dagger \equiv c_{i\uparrow}^\dagger c_{i\downarrow}^\dagger$ and the hard-core constraint $(b_i^\dagger)^2=0$. These are separated from states with unpaired fermions on any site by a gap of order $|U|$. These higher energy states are projected out by a Schrieffer-Wolff transformation on the fermion Hamiltonian in \eqref{fermHam} to obtain the effective low energy Hamiltonian of the on-site bosons. Upto second order in $t,t_\perp$,
\begin{align}
H=&H_{on{\text -}rung}+H_{inter{\text -}rung}-2\mu \sum_{i} n^{(b)}_i \label{Heff}.
\end{align}
Here, $H_{on{\text -}rung}$ and $H_{inter{\text -}rung}$ describe pair hopping and repulsive interaction between the nearest neighbour sites on a rung and on the same layer in adjacent rungs, respectively, and $n^{(b)}_{i}=b_i^\dagger b_i$ is the boson density operator. 

$H_{on{\text -}rung}$ has the ground state 
\begin{align}
|+ \rangle \equiv \frac{1}{\sqrt{2}} \left( b^\dagger_{1} + b^\dagger_{2} \right)|0\rangle \label{bosonGS}
\end{align}
at each rung with energy $-J_\perp=-4t_\perp^2/|U|$ (c.f. Eq. \ref{atomicGS}). Here $\nu=1,2$ indexes the two sites on the upper and lower layer of the rung. The other three eigenstates of $H_{on{\text -}rung}$ at each rung are the degenerate states
\begin{align}
&|0\rangle,\no\\
&|d \rangle \equiv b^\dagger_{1} b^\dagger_{2} |0\rangle ,\no\\
&|- \rangle \equiv \frac{1}{\sqrt{2}} \left( b^\dagger_{1} - b^\dagger_{2} \right) |0\rangle \label{bosonexc}
\end{align}
with zero energy. The insulating ground state at half-filling may be interpreted as a Mott insulator of one symmetrically occupied boson per rung. 

\begin{figure}
\centering
\includegraphics[width=0.45\textwidth]{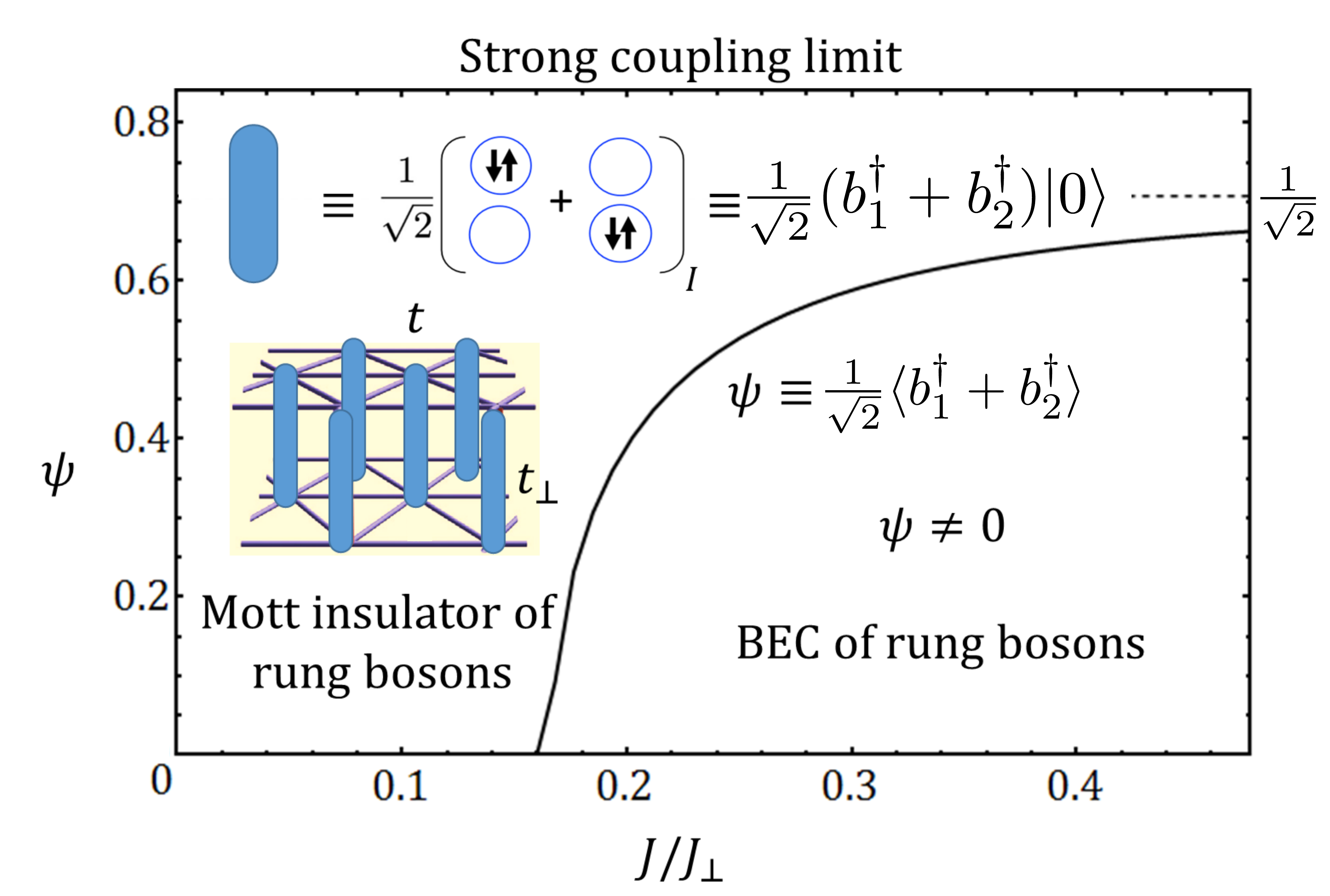}
\caption{Bosonic mean field theory in the strong coupling limit at $T=0$: Fermionic degrees of freedom are gapped out and the low energy modes are hard-core bosons, created by $b^\dagger_{I1(2)}$ on the 1st (2nd) site of the $I$-th rung.
When $t\ll t_\perp$, these bosons delocalise across the rung and at half-filling of the lattice, we have a Mott insulator of one boson per rung. With increasing in-plane hopping, these rung bosons condense into a BEC superfluid with the uniform order parameter $\psi=\langle \frac{1}{\sqrt{2}} \left(b_{I1}+b_{I2}\right)\rangle_{_I}$, where subscript denotes an average over rungs in addition to thermal average. The evolution of the order parameter is shown as a function of $J/J_\perp=(t/t_\perp)^2$.
}
\label{f:strongcoupling}
\end{figure}

$H_{inter{\text -}rung}$ is approximated by an on-rung coupling to a mean field order parameter $\psi= \langle b_1 + b_2 \rangle/\sqrt{2}$ at each rung following Ref.~\cite{fisherBosonLocalizationSuperfluidinsulator1989,sheshadriSuperfluidInsulatingPhases1993}. This gives an on-rung mean-field Hamiltonian (details in Appendix~\ref{a:boson}) represented in the basis of states $|+\rangle, |0\rangle, |d\rangle, |-\rangle$ by the matrix

\begin{align}
H_{MF}\equiv 
\begin{pmatrix}
- J_\perp  & -3J \psi & -3J \psi & 0 \\
-3J \psi 			 & 0 & 0 & 0 \\
-3J \psi & 0 & 0  & 0 \\
0 & 0 & 0 & 0
\end{pmatrix} \label{eqHMF}
\end{align}
where $J=4t^2/|U|$. In the ``strongly interacting" limit of $J \ll J_\perp$, the ground state is a Mott insulator of one symmetric boson per rung : $\prod_{_I} |+\rangle_{_I}$, where $I$ is a rung index, and the self-consistent value of the order parameter $\psi$ is zero. In the opposite limit of $J\gg J_\perp$, the Hamiltonian is identical to a three-site tight-binding problem and has the ground state $\prod_{_I} \left[\frac{1}{2}|0\rangle_{_I}+\frac{1}{\sqrt{2}} |+\rangle_{_I} + \frac{1}{2} |d\rangle_{_I}\right]$. In this limit, the order parameter takes its maximum value $\psi=1/\sqrt{2}$
\begin{align}
\psi=\frac{1}{4} \langle + |_{_I} \left( b_{1} + b_{2} \right) | d\rangle_{_I} + \langle 0 |_{_I} \left( b_{1} + b_{2} \right)| +\rangle_{_I} =\frac{1}{\sqrt{2}}.
\end{align}
The transition from Mott insulator to superfluid occurs at a critical $J/J_\perp=0.16$ which corresponds to $t/t_\perp=0.4$ (Fig \ref{f:strongcoupling}). 

\section{Conclusion}\label{disc}

In this paper, we have studied the superconductor-insulator transition and the BCS-BEC crossover in a simple disorder-free model of lattice fermions. This allows us to distill essential features of both phenomena in absence of complications from competing orders.

Whenever there is a direct transition between an insulator and a superconductor, the two-particle gap must close. Thus if single-particle excitations remain gapped, the insulator just before the transition must be a Bose Insulator, whose low energy excitations are necessarily bosonic. This very general feature of direct SIT is demonstrated from QMC data in Fig.~\ref{f:Gapstory}.

On the superconducting side, the critical energy scale which goes soft at the transition is the superfluid stiffness. Again, if single particle excitations are gapped at the transition, the superconductor just after the transition is a BEC-type superconductor whose phase stiffness $D_s<E_{g1}$ the pair-breaking energy scale. Thus, in this regime close to the transition, it is necessarily phase fluctuations that are the prime determinant of the critical temperature $T_c$. 

A natural corollary is that close to the SIT on the superconducting side, there exists a pairing pseudogap regime that exists over a range of temperature $E_{g1}\sim T^*>T>T_c\sim D_s$ where thermal phase fluctuations have destroyed superconductivity but single-particle excitations are still gapped. 

The BEC superconductor may undergo a further crossover to a BCS type superconductor with a well-defined minimum gap contour in momentum space. We discuss and demonstrate several criteria that may be used to distinguish the two regimes of the crossover, and emphasize that these need not converge to a single ``crossover value'' of the tuning parameter, beyond which characteristic features of the limiting regime are observed. 
Our results and discussions are relevant for recent discussions of the pseudogap phase above $T_c$ in the iron-based superconductors where $\Delta\sim \ef$~\cite{yangBCSlikeCriticalFluctuations2017,hanaguriQuantumVortexCore2019}.

The route from band insulator to BCS superconductor therefore necessarily goes through two intermediate crossover regimes of Bose insulator and BEC superconductor as indicated in Fig.~\ref{f:SchematicEvol}. We showed in Fig.~\ref{f:Akwfig2} how the single-particle spectral function evolves smoothly between these two well-understood limits. This supports the intuition that in a direct SIT, the single-particle degrees of freedom are not important for the transition. 

Going forward, the formalism we have developed is relevant for layered materials such as transition metal dichalcogenides with multi-bands in which the richness of the phases and phase transitions we have outlined can be explored. 

\section{Acknowledgments}
We thank Richard Scalettar for invaluable help in this project. T. H. and M.R. acknowledge support from NSF DMR-1410364. N.T. acknowledges partial support from NSF-DMR 1309461 and from DOE grant DE-FG02-07ER46423.

\bibliography{thesisReferences4}

\appendix

\section{Spectral function in the two-site limit}\label{a:atomic}

In this section, we derive the exact analytical expressions for the single-particle spectral function and the two-particle Green's function in the atomic limit ($t=0$). Exact diagonalization of the two-site Hamiltonian in Eq. \ref{atomicHam} yields the ground state 
\begin{align}
|\psi_0\rangle = &\frac{1}{\sqrt{2}}\big[ \cos \frac{\theta}{2} \left( c_{1\uparrow}^\dagger c_{1\downarrow}^\dagger + c_{2\uparrow}^\dagger c_{2\downarrow}^\dagger \right) 
\no\\ &\quad+ \sin \frac{\theta}{2}\left( c_{1\uparrow}^\dagger c_{2\downarrow}^\dagger + c_{2\uparrow}^\dagger c_{1\downarrow}^\dagger \right) \big] |0\rangle \\
= &\left[ \mathcal{N}_+ c_{+\uparrow}^\dagger c_{+\downarrow}^\dagger + \mathcal{N}_- c_{-\uparrow}^\dagger c_{-\downarrow}^\dagger \right] |0\rangle \label{atomicGS}
\end{align}
with energy $E=-E_0\equiv-\sqrt{U^2 + 16t_\perp^2}/2$ and two fermions on a rung. Here, $c_{\pm\sigma}=(c_{1\sigma} \pm c_{2\sigma})/\sqrt{2}$, $\mathcal{N}_+= \cos(\pi/4- \theta/2)$, $\mathcal{N}_-= \sin(\pi/4- \theta/2)$ and $\theta=\tan^{-1} (4t_\perp/|U|)$. The single-particle excited states are 
\begin{align}
\mathcal{N}_-^{-1} c_{+\sigma}^\dagger |\psi_0\rangle = c_{+\sigma}^\dagger c_{-\uparrow}^\dagger c_{-,\downarrow}^\dagger |0\rangle
\end{align}
with energy $E=t_\perp$ and
\begin{align}
\mathcal{N}_+^{-1} c_{-\sigma}^\dagger |\psi_0\rangle = c_{-\sigma}^\dagger c_{+\uparrow}^\dagger c_{+\downarrow}^\dagger |0\rangle
\end{align}
with energy $E=-t_\perp$. The corresponding states with one hole per rung are
\begin{align}
\mathcal{N}_+^{-1} c_{+\sigma} |\psi_0\rangle = (-1)^\sigma c_{+,-\sigma}^\dagger |0\rangle
\end{align}
with energy $E=-t_\perp$ and
\begin{align}
\mathcal{N}_-^{-1} c_{-\sigma} |\psi_0\rangle = (-1)^\sigma c_{-,-\sigma}^\dagger |0\rangle
\end{align}
with energy $E=t_\perp$. The single particle Green's function in imaginary time $G_{\alpha}(\tau)=\langle c_{\alpha\sigma}(\tau) c_{\alpha\sigma}^\dagger(0) \rangle$ for $0<\tau<\beta$ and $\alpha=\pm$ is given in terms of the spectral function $A_{\alpha}(\omega)$ 
\begin{align}
G_{\alpha }(\tau) = \int_{-\infty}^\infty d\omega \frac{e^{-\omega\tau}}{1+ e^{-\beta\omega}} A_{\alpha}(\omega) 
\end{align}
which in turn is defined in terms of the retarded Green's function as $A_{\alpha}(\omega)=-{\rm Im } G^R_{\alpha}(\omega)/\pi$ with
\begin{align}
&G^R_{\alpha}(\omega) = \sum_{mn} \frac{e^{-\beta E_m}}{\mathcal{Z}} \times \Bigg( \frac{\langle\psi_m | c_{\alpha\sigma} | \psi_n \rangle \langle \psi_n | c^\dagger_{\alpha\sigma} |\psi_m \rangle}{\omega +\ i 0^+ -(E_n-E_m)} 
\no\\ &\qquad\qquad\qquad\qquad+ \frac{\langle\psi_m | c_{\alpha\sigma}^\dagger | \psi_n \rangle \langle \psi_n | c_{\alpha\sigma} |\psi_m \rangle}{\omega +\ i 0^+ +(E_n-E_m)} \Bigg).
\end{align}

Here, $|\psi_{m} \rangle$ are exact eigenstates of Eq. \ref{atomicHam} with energy $E_{m}$ and $\mathcal{Z}$ is the corresponding partition function. The mixing of the bands in the ground state (Eq. \ref{atomicGS}) results in some probability of exciting a particle or a hole in either band. This leads to the two-pole form of the spectral function in Eq. \ref{atomicAkw}.

In similar fashion, we can obtain the two-particle Green's function $P_{ij}(\tau)=\langle c_{i\downarrow} c_{i\uparrow} (\tau) c_{j\uparrow}^\dagger c_{j\downarrow}^\dagger (0)\rangle = \int^\infty_{-\infty} (\mathrm{d}\omega/\pi) \, {\rm Im } P_{ij}^R(\omega) e^{-\omega\tau}/(1-e^{-\beta\omega}) $. For the on-site component, we obtain
\begin{align}
P^R_{11}(\omega) = \sum_{mn} &\frac{e^{-\beta E_m}}{\mathcal{Z}} \times \Big( \frac{\langle\psi_m | c_{1\downarrow} c_{1\uparrow} | \psi_n \rangle \langle \psi_n | c^\dagger_{1\uparrow} c^\dagger_{1\downarrow} |\psi_m \rangle}{\omega +\ i 0^+ -(E_n-E_m)} 
\no\\ &- \frac{\langle\psi_m | c^\dagger_{1\uparrow} c^\dagger_{1\downarrow} | \psi_n \rangle \langle \psi_n | c_{1\downarrow} c_{1\uparrow}  |\psi_m \rangle}{\omega +\ i 0^+ +(E_n-E_m)} \Big). \label{twopGFRetarded}
\end{align}
The inter-site component $P^R_{12}$ is similarly calculated and yields an identical result because the ground state, the two-particle excited states and the matrix elements between them in Eq. \ref{twopGFRetarded} respect inversion symmetry across the rung $1 \leftrightarrow 2$. We find that at half-filling in the limit $t/t_\perp=0$,
\begin{align}
&P_{11}(\tau)=P_{12}(\tau)=P_{22}(\tau)= 
\no\\&\quad 
\frac{1}{2}  \frac{\cos^2 (\theta/2)}{1-e^{-\beta E_{g2}}} \left[ e^{-E_{g2}\tau} + e^{-E_{g2}(\beta-\tau)} \right],
\end{align}
where $E_{g2}=E_0-|U|/2$.

\section{Bosonic Mean field derivation}\label{a:boson}

In this section, we derive the bosonic mean-field Hamiltonian in Eq.~\ref{eqHMF} from the low energy bosonic Hamiltonian in the limit of strong coupling $|U|\gg t,t_\perp$ given by (c.f. Eq. \ref{Heff})
\begin{align}
H=&H_{on{\text -}rung}+H_{inter{\text -}rung}-2\mu \sum_{i} n^{(b)}_i.
\end{align}
where each site $i$ is henceforth indexed by the rung index $I$ and the layer index $\nu=1,2$. Here
\begin{align}
H_{on{\text -}rung}=&-\frac{J_\perp}{2} \sum_I \left(b^\dagger_{I1} b^\phdag_{I2} + h.c. \right) 
\no\\&
+J_\perp \sum_I \left( n^{(b)}_{I1} n^{(b)}_{I2} -\frac{n^{(b)}_{I1} + n^{(b)}_{I2}}{2} \right) 
\end{align}
describes the on-rung pair hopping and repulsion, with $J_\perp=4t_\perp^2/|U|$ and $n^{(b)}_{I\nu}=c^\dagger_{I\nu}c_{I\nu}$, and
\begin{align}
H_{inter-rung}=&-\frac{J}{2} \sum_{\langle IJ \rangle \nu=1,2} \left(b^\dagger_{I\nu} b_{J\nu} + h.c. \right)
  \no\\&
+J \sum_{\langle IJ \rangle \nu=1,2} \left( n^{(b)}_{I\nu} n^{(b)}_{J\nu} -\frac{n^{(b)}_{I\nu} + n^{(b)}_{J\nu}}{2} \right) \label{Hinter}
\end{align}
describes the inter-rung in-plane pair hopping and NN repulsion, with $J=4t^2/|U|$. It is useful to rewrite the inter-rung hopping in terms of the operators that commute with $H_{on-rung}$
\begin{align}
-\frac{J}{2} \sum_{\langle IJ \rangle \nu=1,2} \left(b^\dagger_{I\nu} b^\phdag_{J\nu} + h.c. \right)=
-\frac{J}{2} \sum_{\langle IJ \rangle \alpha=\pm} \left(b^\dagger_{I\alpha} b^\phdag_{J\alpha} + h.c. \right)
\end{align}
where $b_{I\alpha}=(b_{I1}+\alpha b_{I2})/\sqrt{2}$. In the superfluid state, the symmetric bosons $b_{I+}$ condense into a coherent state $\psi \equiv \langle b_{I+} \rangle_{_I}$ and NN hopping between rungs is approximated, within mean field theory, as an on-rung coupling to the uniform bosonic field $\psi$
\begin{align}
&-\frac{J}{2} \sum_{\langle IJ \rangle \alpha} \left(b^\dagger_{I\alpha} b_{J\alpha} + h.c. \right) 
\no\\&
\approx - \frac{J z}{2} \sum_{I} \left[ \left( b^\dagger_{I+} + b_{I+}\right) \psi - \psi^2 \right]
\end{align}
where $z=6$ is the in-plane coordination number. 

The second term in Eq. \eqref{Hinter} which corresponds to in-plane NN repulsion, is approximated by a Hartree shift in the chemical potential, which gives
\begin{align}
&J \sum_{\langle IJ \rangle \nu} \left( n^{(b)}_{I\nu} n^{(b)}_{J\nu} -\frac{n^{(b)}_{I\nu} + n^{(b)}_{J\nu}}{2} \right) 
\no\\&
\approx J \sum_{\langle IJ \rangle \nu} \left[ \left( \bar{n} - \frac{1}{2} \right) \left( n^{(b)}_{I\nu} + n^{(b)}_{J\nu} \right) -\bar{n}^2 \right]
\end{align}
where $\bar{n} \equiv \sum_{I\nu}\langle n^{(b)}_{I\nu} \rangle/N $. The shift in the chemical potential is zero at half-filling and the only effect of the in-plane NN repulsion is a constant shift in the energy which we ignore. We thus arrive at the on-rung mean field Hamiltonian 
\begin{align}
H_{MF}= H_{intra} - & 3J \left( b^\dagger_{+} + b_{+} \right)\psi 
 \no\\&
+ 3J \psi^2  - 2\mu ( n^{(b)}_{1} + n^{(b)}_{2} ) \label{HMF}
\end{align}
where $H_{intra}=-J_\perp |+\rangle_i \langle +|_i $ encapsulates the on-rung terms that we have treated exactly in Eq. \ref{bosonGS},\ref{bosonexc}. Observing that this Hamiltonian is clearly particle hole symmetric, we set $\mu=0$ to restrict our analysis to half-filling. 

\section{Analysis of single-particle and two-particle Green's functions}\label{a:gapextraction}

In this section, we provide details on the analysis of the imaginary-time Green's functions and describe how we extract the quasiparticle energies and weights from the MC data (without analytic continuation). We also demonstrate how we extract the two-particle excitation gap from the two-particle Green's function.

The Green's function $G(\mathbf{k},\tau)=\langle c_{\veck}(\tau) c^\dagger_{\veck}(0) \rangle$ (with spin indices suppressed) is evaluated in DQMC. For $0<\tau<\beta$, this is related to the spectral function $A(\veck,\omega)$ by
\begin{align}
G(\veck,\tau)=\int_{-\infty}^\infty d\omega	\frac{e^{-\omega \tau}}{1+e^{-\beta\omega}} A(\veck,\omega).
\end{align}
It is easy to check that the sum rule on the spectral function $\int_{-\infty}^\infty \mathrm{d}\omega \, A(\veck,\omega) =1$ implies that $G(\veck,0^+)+G(\veck,\beta^-)=1$. This allows us to understand the two limits of Green's function as $G(\veck,0^+)=\langle c_{\veck}(0^+) c^\dagger_{\veck}(0) \rangle \to 1-n_k$ and $G(\veck,\beta^-) \to n_k$. The quasiparticle dispersion $E_\alpha(\veck)$ is given by the positions of poles in the spectral function $A(\veck,\omega)$ at $T=0$ (here $\alpha$ is the band and/or particle-hole label). The contribution of one quasiparticle pole $A_{QP}(\omega)=\delta(\omega-E_\alpha)$ to the Green's function is $e^{-E_\alpha \tau}/(1+e^{-\beta E_\alpha})\approx e^{-E_\alpha \tau}$ when $\beta E_\alpha \gg 0$ and $e^{-E_\alpha \tau}/(1+e^{-\beta E_\alpha}) \approx e^{E_\alpha (\beta-\tau)}$ when $\beta E_\alpha \ll 0$. 
We can thus extract particle and hole quasiparticle energies directly from the imaginary time ($\tau$) dependence of the Green's function. We fit the DQMC data to the form of $G(\veck,\tau) = u_\veck^2 e^{-E_k^p \tau} + v_\veck^2 e^{-E_k^h (\beta-\tau)}$ corresponding to a two-pole spectral function $A(\veck,\omega)=u_\veck^2 \delta(\omega-E_k^p) + v_\veck^2 \delta(\omega+E_k^h)$. This is demonstrated in Fig. \ref{f:Gapstory}(a) for $t/t_\perp=0.2$ (insulator), $\veck=(0,0,\pi)$ (upper band bottom) at $\beta=12$ and half-filling. The black curve has the form discussed above with $u_\veck, v_\veck, E_\veck^p, E_\veck^h$ as fitting parameters.

\begin{figure}
  \centering
  \includegraphics[width=0.45\textwidth]{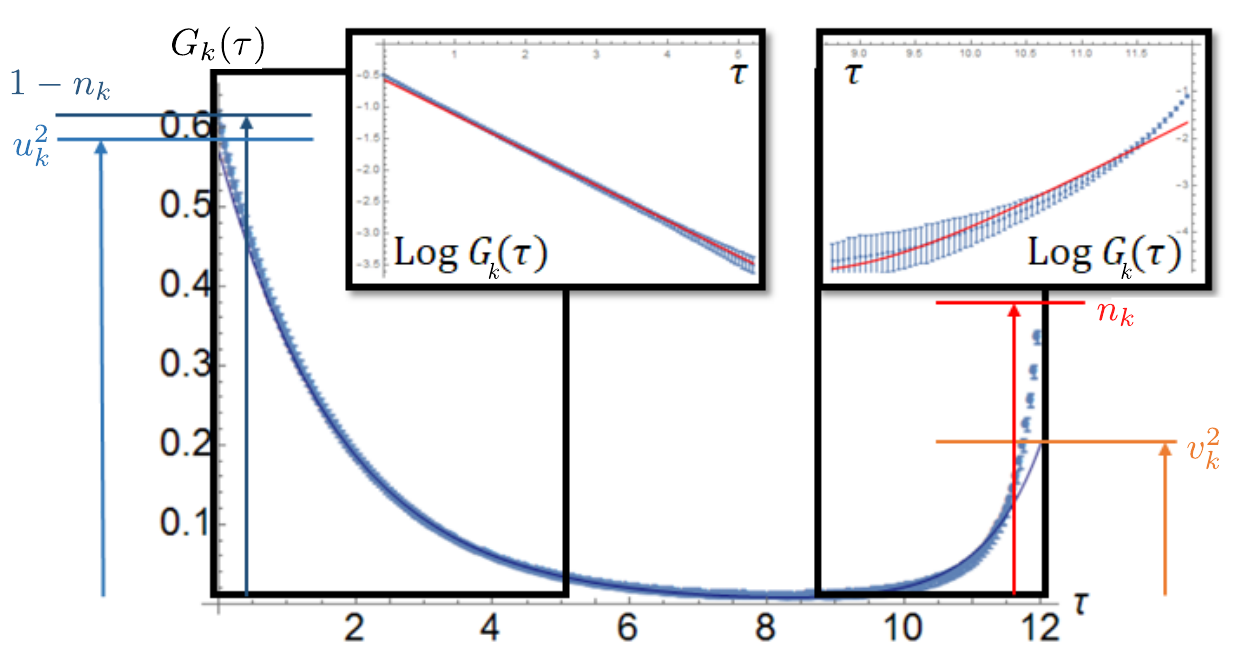}
  \caption{Representative demonstration of the estimation of the quasiparticle weights from the imaginary time dependence of the one-particle Green's function. The difference between the asymptotic value of the fit and the observed value of the $G(k,\tau\rightarrow0,\beta)$ limits gives the intgrated incoherent spectral weight at negative and positive energies respectively.}
  \label{f:QPweight}
\end{figure}

In general, the spectral function has, in addition to the quasiparticle poles, some incoherent weight arising from scattering to other excited states $A(\veck, \omega)=A_{inc}(\veck, \omega) + \sum_\alpha Z_\nu(\veck) \delta(\omega -E_\alpha(\veck)$. Restricting our arguments to positive energies without loss of generality, we note that if this incoherent weight is predominantly at $\omega>E_\alpha$, then its contribution to the Green's function dies off faster than $e^{-E_\alpha \tau}$. The fit then deviates from the data only near $\tau \rightarrow 0$ and the slope of $\log G(\tau)$ at large $\tau$ gives the quasiparticle energy. The deviation of the spectral weights $u_k^2+v_k^2$ from 1 can be used to estimate $\int_{-\infty}^\infty \mathrm{d} \omega \, A_{inc}(\veck,\omega)$. Conversely, if there is significant incoherent weight at $\omega < E_\alpha$, we expect the Green's function to show a systematic deviation from the best fit to $e^{-E_\alpha \tau}$ at $\tau \gg E_\alpha^{-1}$. In this case, fitting the large $\tau$ data then provides a rough estimate of the gap in the single particle spectral function. In our analysis, we find systematic deviations only at $\tau\approx 0, \beta$, and from the fit parameters $u_k^2$ and $v_k^2$, we estimate that atleast $60\%$ of the spectral weight is always contained in the quasiparticle poles and the rest is at higher energies. 
In Fig. \ref{f:QPweight}, we demonstrate the estimation of quasiparticle weight from the imaginary time Green's function close to the SIT, for $t/t_\perp=0.3$.

\begin{figure}
  \centering
  \vspace{10pt}
  \includegraphics[width=0.45\textwidth]{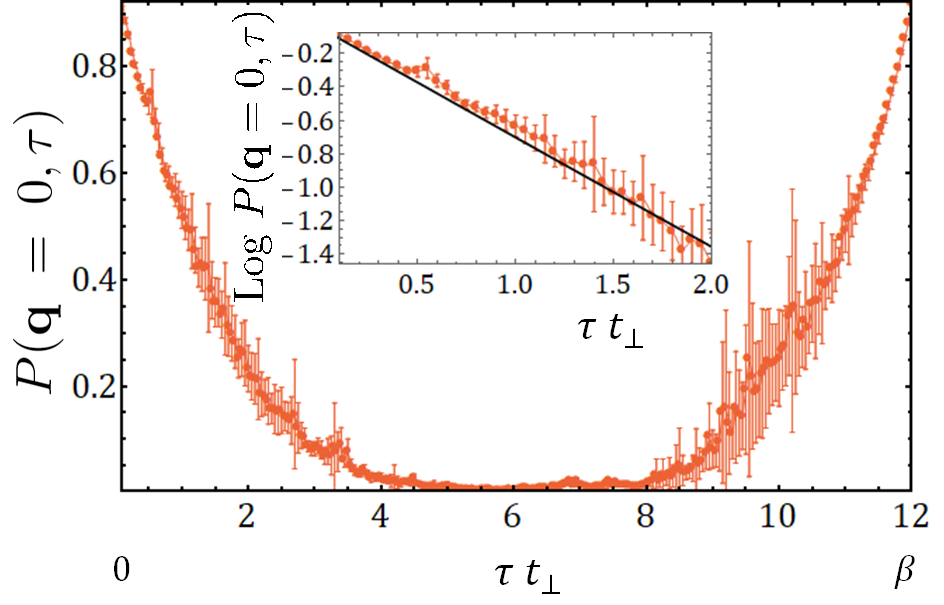}
  \caption{Representative demonstration of the estimation of the two-particle gap from the imaginary time dependence of the two-particle Green's function $P(\mathbf{q}=0,\tau)=N^{-2} \sum_{k,k'} \langle c_{k\uparrow}^\dagger c_{-k\downarrow}^\dagger (\tau) c_{-k\downarrow} c_{k\uparrow} (0) \rangle$. We fit the data only for $\tau \approx 0,\beta$ since the data near $\tau=\beta/2$ is known to be unreliable due to long autocorrelation times. As a result, the energy scale obtained from the fit is the position of the peak in the two-particle spectral function rather than the true spectral gap. MC data at $t=0.2t_\perp$ on a $6 \times6$ bilayer.}
  \label{f:twogap}
\end{figure}

For extracting the two-particle gap, we fit the two particle Green's function in imaginary time $P(\mathbf{q}=0,\tau)=N^{-2}\sum_{k,k^{'}} \langle c_{-k\downarrow} c_{k\uparrow} (\tau) c_{k^{'}\uparrow}^\dagger c_{-k^{'}\downarrow}^\dagger (0) \rangle$ which is given in terms of the retarded two-particle propagator $P^R(\mathbf{q}=0,\omega)$ by 
\begin{align}
P(\tau)=\int \frac{d\omega}{\pi} \frac{e^{-\omega\tau}}{1-e^{-\beta\omega}} {\rm Im} P^R(\omega)
\end{align}
At temperatures small compared to $\Omega$, the lowest energy scale at which there is any structure in ${\rm Im } P^R(\omega)$, the rate of decay of $P(\tau)$ at $\tau\rightarrow 0(\beta)$ is sensitive to the frequency scale at which the two particle spectral function $(1/\pi) {\rm Im } P^R(\omega)$ is peaked for $\omega>0 (\omega<0)$ corresponding to particle (hole) pairs.
\begin{align}
-\left.\frac{\partial \log P(\tau)}{\partial \tau}\right\rvert_{\tau=0^+} \approx \frac{\int_\Omega^\infty \omega {\rm Im } P^R(\omega)d\omega }{\int_\Omega^\infty {\rm Im } P^R(\omega) d\omega } \no \\
-\left.\frac{\partial \log P(\tau)}{\partial \tau}\right\rvert_{\tau=\beta^-} \approx \frac{\int_{-\infty}^{-\Omega} \omega {\rm Im } P^R(\omega)d\omega }{\int_{-\infty}^{-\Omega} {\rm Im } P^R(\omega) d\omega }
\end{align}
The reason we extract the two-particle gap from the small $\tau$ data is that we find that the two-particle Green's function at large $\tau$ takes extremely long to converge in DQMC and has a long autocorrelation time. This restricts the number of independent measurements and results in unreliable data. For small $\tau$, the data is reliable with much fewer measurement sweeps, and it is possible to take a large number of independent measurements to obtain statistically exact measures of $P(\tau)$.
A natural consequence of the separation of energy scales governing phase fluctuations and fermionic excitations is the existence of a pairing pseudogap regime in a temperature range where thermal fluctuations destroy the superconducting order but the single-particle spectrum remains gapped. The pairing pseudogap is ubiqitous in strongly-correlated materials where they are mired in complications from spatial inhomogeneity and competing orders. Here, we are able to extract clear intuition and concrete observables in QMC simulations in this enigmatic regime, without any of these complications. 

\section{Single particle spectral function}\label{Akw}

In this section, we discuss the evolution of the spectral function across the SIT and the BCS-BEC crossover (Fig.~\ref{f:Akwfig2}).
We estimate the position of quasiparticle poles in  the single particle spectral function $A(k,\omega)$ and their quasiparticle weights, using the techniques demonstrated in Appendix~\ref{a:gapextraction}.
We find that the quasiparticle spectrum evolves smoothly from a Fermi insulator to a BCS superconductor and looks remarkably similar for the Bose insulator and the BEC superconductor on either side of the SIT. We also outline a new technique to identify the BCS-BEC crossover in the superconductor by studying the ratio of the particle and hole spectral weights at the band edges, and discuss its merits and demerits.

\begin{figure*}[htb!]
  \centering
  \includegraphics[width=0.8\textwidth]{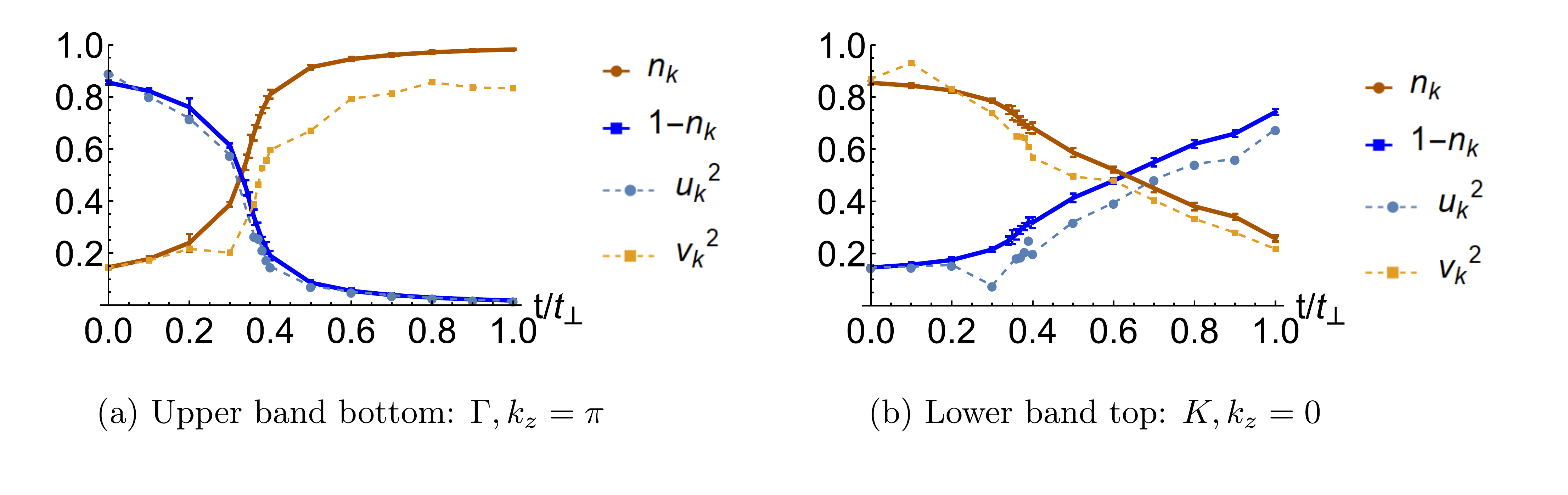}
  \caption{BEC-BCS crossover on the two bands as identified by the ratio of the quasiparticle weights. Dashed lines represent estimates of the quasiparticle weights $Z_k$ in the quasiparticle poles, obtained by fitting $G(k,\tau)$. Solid lines represent integrated spectral weights over positive or negative energies. The crossover is identified by the crossing of these lines. The solid curves are guaranteed to cross at $1/2$ by the sum rule $\int d\omega A(k,\omega) =1$. The crossover can thus be identified from the equal-time measurement $1-n_k=\langle c_k c_k^\dagger \rangle = G(k,0^+)$ alone. (a) Crossover on the upper band as identified by the quasiparticle weights at the $\Gamma$ point. This roughly coincides with the crossover as identified by the topology of the min gap locus. (b) Crossover on the lower band as identified by the quasiparticle weights at the $K$ point. This crossover is not observable from the topology of the min gap locus within the range of parameters available, possibly due to the discreteness of the k-space lattice. MC data on a $6\times 6$ bilayer at $\beta t_\perp =12,|U|=4t_\perp,\Delta\tau=0.05$. 
}\label{f:ukvkcrossover}
\end{figure*}

A prominent feature of the BCS superconducting spectral function is the doubling of the quasiparticle poles due to particle-hole mixing, that distinguishes it from the spectral function of a band insulator. It is interesting to note that this doubling does not appear abruptly at the SIT, but emerges smoothly as interactions are turned on in the insulator. In presence of interactions, the ground state has partial occupation of both bands, due to the presence of pair hopping terms like $-|U| c_{k+\uparrow}^\dagger c_{-k+\downarrow}^\dagger c_{-k-\downarrow} c_{k-\uparrow}$ in the Hamiltonian, where $+(-)$ indicates the $k_z=0(\pi)$ band. This leads to some probability of creating either a particle-like ($E>0$) or a hole-like ($E<0$) excitation on either band. This doubling of the quasiparticle poles is a numerical signature of preformed pairs and incipient superconductivity in an insulator. Although the global $U(1)$ symmetry is not broken, fermion number is no longer separately conserved on each band in the presence of interactions, leading to the observed doubling of quasiparticle poles in (Fig. \ref{f:Akwfig2}(b)).

In the superconducting state, the particle and hole quasiparticle weights on each band are momentum dependent. Within BCS MFT, the momenta inside of the Fermi surface for which the band energy $\epsilon_k < \mu$ have predominantly hole-like excitations $|u_k|^2 < |v_k|^2$ and vice versa. At the band extrema, the ratio of the quasiparticle weights $|u_k^2|/|v_k^2|$ crosses $1/2$ at the BCS-BEC crossover. This, then, is an independent way of characterising the crossover from BEC to BCS physics. Outside of MFT, this may not coincide with the point when the min gap locus changes from point to contour if there is some incoherent spectral weight. This procedure is particularly useful on finite sized lattices, where the granularity of the momentum space makes the topology of the min-gap locus hard to evaluate.


Alternatively, the integrated spectral weight at positive and negative energies ($1-n_k$ and $n_k$ respectively) can be read off from the $\tau\rightarrow 0,\beta$ limits of the single particle spectral function (Appendix~\ref{a:gapextraction}). This quantity can also be used to independently identify the crossover regime, in presence of incoherent spectral weight. This is also useful on small lattices, and additionally requires only the equal-time Green's function (which is much easier to calculate in QMC).

In Fig. \ref{f:ukvkcrossover}, we demonstrate the use of these two techniques in identifying the BEC-BCS crossover in this model. We emphasize that this is independent of the topology of the min gap locus: these methods are able to identify crossovers on both bands, whereas the min gap criterion does not show a change in min gap locus topology on the lower band ($k_z=0$) in the parameter regime explored (upto $t=2t_\perp$). Fig. \ref{f:ukvkcrossover} clearly identifies three distinct regimes of the superconductor, one in which both bands are BEC-like, an intermediate regime where one band is BCS-like and the other is still BEC-like, and a third in which both are BCS-like.

Another criterion for identifying the crossover is the lowest energy scale that destroys superconductivity: the one-particle gap in the BCS regime and the superfluid stiffness in the BEC regime. Unlike the previous criteria which are band selective, this is a global criterion that takes all bands into account. In the regime of parameters we have explored, the superconductor is always in the BEC regime according to this criterion.

\end{document}